\documentclass[manuscript]{acmart}

\definecolor{AbiOrange}{RGB}{251,212,180}
\definecolor{TimBlue}{RGB}{219,229,241}

\newcommand{\NONE}[1]{{%
\leavevmode\color{black}
\textsl{NONE}}}

\newcommand{\ALL}[1]{{%
\leavevmode\color{black}
\textsl{ALL}}}

\newcommand{\STT}[1]{{%
\leavevmode\color{black}
\textsl{STT}}} 
\newcommand{\STTlong}[1]{{%
\leavevmode\color{black}
Scores Through-Time}}

\newcommand{\OTB}[1]{{%
\leavevmode\color{black}
\textsl{OTB}}} 
\newcommand{\OTBlong}[1]{{%
\leavevmode\color{black}
Scores On-the-Board}}

\newcommand{\BTW}[1]{{%
\leavevmode\color{black}
\textsl{BTW}}} 
\newcommand{\BTWlong}[1]{{%
\leavevmode\color{black}
Scores Best-to-Worst}}

\newcommand{\REDACT}[1]{$\Box\Box\Box$} 

\newcommand{\redactCollege}[1]{[a U.S. University]}  

\newcommand{\quotateInset}[4]{%
\vspace{-3pt}%
\begin{quote}%
     \leftskip-10pt
     \rightskip-15pt
\textbf{#1}: \emph{``#4''}\end{quote}%
\vspace{3pt}}

\newcounter{boldifyCounter}
\newcounter{fixmeSectionCounter}
\newcounter{fixmeTotalCounter}
\makeatletter
\@addtoreset{fixmeSectionCounter}{section}
\@addtoreset{fixmeSectionCounter}{subsection}
\@addtoreset{boldifyCounter}{section}
\@addtoreset{boldifyCounter}{subsection}
\makeatother

\newcommand{\boldify}[1]{}
\ifdefined\boldifyON
	\renewcommand{\boldify}[1]{
        \par\noindent
		\stepcounter{boldifyCounter}
		\textbf{{\color{green}**}
		~\arabic{section}.\arabic{subsection}.\arabic{boldifyCounter}
		: #1} 
	}
\fi 

\newcommand{\reportOnFIXME}{%
    \newcount\iterCounter
    \iterCounter=1
    \newcount\endCounter
    \endCounter=\totvalue{fixmeTotalCounter}
    \advance \endCounter +1
    There are 
    {\color{red}\total{fixmeTotalCounter}} 
    FIX\_ME\\
    links:
    \loop
        \hyperlink{fixTag\the\iterCounter}{\#\the\iterCounter}
        \advance \iterCounter +1
    \ifnum \iterCounter < \endCounter
    \repeat
}

\newcommand{\FIXME}[1]{} 
\ifdefined\fixmeON
	\renewcommand{\FIXME}[1]{\par\noindent
		\stepcounter{fixmeSectionCounter}\stepcounter{fixmeTotalCounter}
		{\color{red}\fbox{\color{black}
			\parbox{.99\linewidth}{
				\textbf{\hypertarget{fixTag\thefixmeTotalCounter}{FIXME}	\arabic{section}.\arabic{subsection}.
        		\arabic{fixmeSectionCounter} (\color{red}
        		\#\arabic{fixmeTotalCounter}):} #1}}
        }
	}
\fi 

\newcommand{\FIXED}[1]{}
\ifdefined\fixmeON
	\renewcommand{\FIXED}[1]{\par\noindent%
		{\color{black}\fbox{\color{black}%
			\parbox{.99\columnwidth}{%
				\color{blue}#1}}%
        }
	}
\fi 

\newcommand{\draftStatus}[1]{}
\ifdefined\draftStatusON
	\renewcommand{\draftStatus}[1]{
        \hfill **#1
	}
\fi 

\usepackage{calc, enumitem} 
\usepackage{soul}
\usepackage{xcolor, colortbl}
\usepackage{multirow}
\usepackage{rotating}
\usepackage{subcaption}
\usepackage{stfloats}
\usepackage{wasysym} 
\usepackage{totcount}\regtotcounter{fixmeTotalCounter}
\usepackage{datetime2} 

\copyrightyear{2025} 
\acmYear{2025} 
\setcopyright{acmcopyright}\acmConference[TOSEM]{ACM Transactions on Software Engineering and Methodology}{}{}
\acmBooktitle{ACM Transactions on Software Engineering and Methodology}
\acmPrice{15.00}
\acmDOI{10.1145/3490099.3511115}
\acmISBN{978-1-4503-9144-3/22/03}

\begin{document}

\title{How to Measure Human-AI Prediction Accuracy in Explainable AI Systems}



\author{Sujay Koujalgi}
\email{snk5290@psu.edu}
\affiliation{%
  \institution{%
  Penn State University}
  \streetaddress{Westgate Building}
  \city{University Park}
  \state{PA}
  \country{USA}
  \postcode{16802}
}

\author{Andrew Anderson}
\email{anderan2@oregonstate.edu}
\affiliation{%
  \institution{%
  Oregon State University}
  \streetaddress{1148 Kelley Engineering Center}
  \city{Corvallis}
  \state{OR}
  \country{USA}
  \postcode{97331-5501}
}

\author{Iyadunni Adenuga}
\email{ija5027@psu.edu}
\affiliation{%
  \institution{%
  Penn State University}
  \streetaddress{Westgate Building}
  \city{University Park}
  \state{PA}
  \country{USA}
  \postcode{16802}
}

\author{Shikha Soneji}
\email{sxs7000@psu.edu}
\affiliation{%
  \institution{%
  Penn State University}
  \streetaddress{Westgate Building}
  \city{University Park}
  \state{PA}
  \country{USA}
  \postcode{16802}
}

\author{Rupika Dikkala}
\email{dikkalar@oregonstate.edu}
\affiliation{%
    \institution{%
    Oregon State University
    }
    \streetaddress{1148 Kelley Engineering Center}
    \city{Corvallis}
    \state{OR}
    \country{USA}
    \postcode{97331-5501}
}

\author{Teresita Guzman Nader}
\email{guzmannt@oregonstate.edu}
\affiliation{%
    \institution{%
    Oregon State University
    }
    \streetaddress{1148 Kelley Engineering Center}
    \city{Corvallis}
    \state{OR}
    \country{USA}
    \postcode{97331-5501}
}

\author{Leo Soccio}
\email{lss5473@psu.edu@psu.edu}
\affiliation{%
  \institution{%
  Penn State University}
  \streetaddress{Westgate Building}
  \city{University Park}
  \state{PA}
  \country{USA}
  \postcode{16802}
}

\author{Sourav Panda}
\email{sbp5911@psu.edu}
\affiliation{%
  \institution{%
  Penn State University}
  \streetaddress{Westgate Building}
  \city{University Park}
  \state{PA}
  \country{USA}
  \postcode{16802}
}

\author{Rupak Kumar Das}
\email{rjd6099@psu.edu}
\affiliation{%
  \institution{%
  Penn State University}
  \streetaddress{Westgate Building}
  \city{University Park}
  \state{PA}
  \country{USA}
  \postcode{16802}
}

\author{Margaret Burnett}
\email{burnett@eecs.oregonstate.edu}
\affiliation{%
  \institution{%
  Oregon State University}
  \streetaddress{1148 Kelley Engineering Center}
  \city{Corvallis}
  \state{OR}
  \country{USA}
  \postcode{97331-5501}
}

\author{Jonathan Dodge}
\email{jxd6067@psu.edu}
\affiliation{%
  \institution{%
  Penn State University}
  \streetaddress{Westgate Building}
  \city{University Park}
  \state{PA}
  \country{USA}
  \postcode{16802}
}

\renewcommand{\shortauthors}{Koujalgi et al.}

\begin{abstract}
Assessing an AI system's behavior---particularly in Explainable AI Systems---is sometimes done empirically, by measuring people's abilities to \textit{predict} the agent's next move---but how to perform such measurements? 
In empirical studies with humans, an obvious approach is to frame the task as binary (i.e., prediction is either right or wrong), but this does not scale.
As output spaces increase, so do floor effects, because the ratio of right answers to wrong answers quickly becomes very small.
The crux of the problem is that the binary framing is failing to capture the nuances of the different degrees of ``wrongness''.
To address this, we begin by proposing three mathematical bases upon which to measure ``partial wrongness''.
We then uses these bases to perform two analyses on sequential decision-making domains: the first is an in-lab study with 86 participants and a size-36 action space; the second is an analysis of a size-4 action space.
\FIXME{MMB says: is there anything interesting we can say about our results?}
Other researchers adopting our operationalization of the prediction task and analysis methodology will improve the rigor of user studies conducted with that task, which is particularly important when the domain features a large output space.
\end{abstract}


%

\maketitle

\section{Introduction}\label{sectionIntro}

\boldify{AI needs SE, and it's starting to appear}

When AI began to make its way into mainstream software development, many researchers began calling for software engineering methods suitable for AI systems (SE4AI) (e.g.,~\cite{belani2019RE4AI, hill2016trialsAIdevs, ozkaya2020SE4AI-software, sculley2015technicalDebtML}). 
In response, the software engineering community has begun to create with SE4AI methods and tools, recently surveyed by Martinez-Fernandez et al. and by Giray~\cite{martinez2022SE4AIsurvey, giray2021SE4MLsurvey}.

\boldify{But lots of gaps, especially in the area of transparency -- and it turns out transparency really matters}

Still, gaps remain, especially in SE4AI methods for attributes that are unique to AI-powered systems~\cite{martinez2022SE4AIsurvey}.  
One such attribute is \textit{transparency}.~
\footnote{In this paper, we do not distinguish between transparency and its synonyms understandability, comprehensibility, interpretibility, scrutibility, explanability.} 
In some software, transparency is considered to be so critical, it is a legal requirement.  
For example, U.S. Executive Order 13960 requires ``certain federal agencies'' to abide by nine principles except in cases of national security, two of which are understandability and transparency~\cite{executiveOrder13960}.
Likewise, the European Union's 2023 AI Act introduces specific transparency obligations
~\cite{EU-AI-Act2023}.

%

\boldify{XAI features' goal is transparency -- but how to assess whether the explanations they produce are actually understandable by humans? }

The subarea of eXplainable AI (XAI) exists to fulfill such requirements of transparency and understandability.
However, how to assess whether an XAI feature actually achieves  of this property is not yet well established.

\boldify{Here we come to save the day, using the PREDICTION task.}

This paper aims to shed light on how to assess an XAI system's transparency using several mathematical measures in empirical studies with humans.
Drawing from Hoffman et al.'s~\cite{hoffman2018metrics} suggested of eleven empirical tasks researchers could rely upon to evaluate their explanations (see their Tables~4 and 5), 
we consider several ways how well human participants can \textit{predict} an AI system's next decision.
Using human predictions~\cite{muramatsu2001transparent} is an important way to assess human understanding that has seen widespread use (e.g., \cite{soratana2021human,anderson2020mental,dai2022counterfactual,nourani2021anchoring}).

\FIXME{MMB@anyone: the next two paragraphs make a lot of the same arguments twice. Some merging and cutting needed...}

\boldify{Operationalizing prediction is hard, but here is how it works and is often framed/analyzed.
The obvious thing is binary correct/wrong, but this is very susceptible to floor and ceiling effects.
}

Measuring human performance on predicting AI behaviors is difficult, in part because the obvious binary measure (correct/wrong) produces very little useful information. 
Consider the following example: a self-driving car approaches a construction zone with cones defining the lane instead of painted lines.
One prediction question we might ask is \emph{``Will it hit any cones? Yes or No.''}
This binary framing is fairly common in XAI research (e.g., \cite{soratana2021human,anderson2020mental,dai2022counterfactual,nourani2021anchoring}), but it is very susceptible to the floor and ceiling effects these authors report.
In particular, some decisions are very easy and \textit{most} participants get them correct; while other decisions are more difficult and \textit{no one} gets them correct.
The second situation is particularly dangerous in XAI because as the output space grows, the probability of any particular decision being correct shrinks quite rapidly.

The main contribution of this paper is a method for assessing XAI's transparency property using the empirical Prediction task.
As such, our work improves our ability to empirically study a non-functional requirement: \ul{that an agent's behavior is predictable}.
The current method of binary prediction framing does not account for partially correct predictions.
As a result, binary prediction measures are susceptible to Type 2 errors, which occur \textit{``when we declare no differences or associations between study groups when, in fact, there was''}~\cite{shreffler2023}.
Such Type 2 errors become even more likely as output space size increases.
One way to counteract this trend could be to incorporate domain-specific knowledge, but that would not generalize well.
Some actions \emph{appear} similar, some actions produce similar \emph{outcomes}, but knowing when similarities are merely superficial requires domain knowledge.
Instead, our approach is domain-agnostic: it requests values (and rank-order preferences) from the agent to measure ``partial credit,'' providing an alternative to binary prediction analysis. 
Note that our studies focused on controlling the size of the action space, while many other games have action spaces that are far larger; in an extreme case DeepMind's StarCraft parameterization \textit{``has an average of approximately $10^{26}$ legal actions at every time-step''}~\cite{vinyals2019grandmaster}.
Clearly a binary prediction framing is ill-suited to such an action space.

\FIXME{MMB@anyone: from here on, some of the same arguments keep getting made different ways.  Some should be cut.}

\FIXME{MMB@anyone: the related work in the next parag should instead move to the Background/Related Work seciton.  Why: It's breaking the flow here.}

\boldify{So our core goal is to determine some way to define ``partially correct'' in a robust way for either the continuous case or by discretizing.}

Our core goal is to rigorously define measures for the meaning of ``partially correct'' when predicting an AI output.
Some recent work by Bondi et al.~\cite{bondi2022role} moves past a binary framing by having participants  enter their prediction on a 5-point Likert scale that ranges from \textit{definitely not present} to \textit{definitely present}.
Unfortunately, the higher granularity in that measure comes from the \emph{participants' certitude and not the agent}.
Inspired by directions like Bondi et al.'s, this paper proposes new strategies to operationalize predictions so as to incorporate a well-developed notion of ``partial credit.''

\boldify{The key insight in this paper is to use the outputs of the agent itself to assign partial credit.
From there, we make some analysis methods and show them off (END CONTRIBUTIONS)}

Another, more granular, question we might ask in our situation from earlier is, \emph{``Which cone will it hit, if any?''}
The key insight in this paper is that it is possible to assign partial credit based on the \emph{outputs of the agent itself}.
Based on this insight, we contribute several novel methods to analyze human prediction task data.
Additionally, we offer two different demonstrations of our analysis approach applied to domains with markedly different action space sizes.
Note that whether or not a process yields significance on a particular dataset does not determine the validity of the process.
The data are only to illustrate what the methods reveal.

\boldify{The first is a quantitative study with 86 participants using MNK games}

The first study is a newly conducted in-lab study, recruiting a total of 86 participants.
For this study, we adapted the public source in MNK games provided by Dodge et al.~\cite{dodge2022people} for agents and explanations, due to people's familiarity with such games.
In particular, we used 9-4-4, meaning that the board contained $9 \times 4 = 36$ squares, and 4-in-a-row would be a winning sequence.
In this domain, we asked participants to predict the square (row and column index) that an agent would take in the next move: out of the 36 options.

\boldify{The second is a reanalysis of a previously published study using a simple 4 towers domain}

The second study is a reanalysis of data from a previously published study, by Anderson et al.~\cite{anderson2020mental}.
That study used a custom domain (Four Towers) intended to tightly control the action space to just 4 options.
In that domain, the study's authors asked participants to predict the quadrant that the agent would attack, out of four options (NE, NW, SE, SW).

\boldify{We set out to answer THESE RQs, which are pretty great and you should agree:}

This paper seeks to address the following research questions:
\begin{enumerate}[leftmargin=25pt]

    \item[\textbf{RQ1}] \emph{Analyzing-Prediction-Task}: 
    How can we operationalize analyzing prediction task data in a way that has some degree of independence from domain and model?

    \item[\textbf{RQ2}] \emph{Study2-Predict-Among-36}: 
    How well did participants predict the agent's actions with an action space of size 36 in an MNK game domain?

    \item[\textbf{RQ3}]\emph{Study1-Predict-Among-4}: 
    How well did participants predict the agent's actions with an action space of size 4 in the Four Towers domain?

\end{enumerate}

\textbf{RQ1} is our main RQ; it asks how to solve the described problem with binary prediction.
\textbf{RQ2} and \textbf{RQ3} are vehicles for us to investigate and illustrate the methods derived answering \textbf{RQ1} with two concrete domains, (small) action space sizes, and concrete data sets. 
While we have motivated that this issue is more pressing with very large action spaces, we start with small action spaces where both clarity of explanation and scientific control are more tractable.

\FIXME{MMB to JED: I tried to get this closer to what I thought SE reviewers might like.  IF there's anything I did that you don't like, feel free to revert to the original version.}

\section{Background and Related Work}
\label{sectionBackground}



\FIXME{MMB: just flagging that this section hasn't been finished yet.\\
JED: commented out the above pieces so it would be a better arxiv submission.

Chen 2023
Quality-oriented Testing for Deep Learning Systems:

Kexin Pei, Yinzhi Cao, Junfeng Yang, and Suman Jana. 2017. 
DeepXplore: Automated whitebox testing of deep learning systems.

Aggarwal
Testing framework for black-box AI models: 

Sainyam Galhotra, Yuriy Brun, and Alexandra Meliou. 2017. 
Fairness testing: Testing software for discrimination.

Aniya Aggarwal et al. 2019.
Black box fairness testing of machine learning models.

Model-based hypothesis testing of uncertain software systems:

Hazem Fahmy et al. 2022.
HUDD: A tool to debug DNNs for safety analysis.

Groce
You Are the Only Possible Oracle: Effective Test Selection for End Users of Interactive Machine Learning Systems: 

Improving model-based test generation by model decomposition:

Nicholas Carlini and David Wagner. 2017.
Towards evaluating the robustness of neural networks.

Nicholas Carlini, Anish Athalye, Nicolas Papernot, Wieland Brendel, Jonas Rauber, Dimitris Tsipras, Ian Goodfellow, Aleksander Madry, and Alexey Kurakin. 2019.
On evaluating adversarial robustness.

Wieland Brendel, Jonas Rauber, Matthias Kümmerer, Ivan Ustyuzhaninov, and Matthias Bethge. 2019.
Accurate, reliable and fast robustness evaluation.

Ian Goodfellow, Jonathon Shlens, and Christian Szegedy. 2015. 
Explaining and harnessing adversarial examples.

Aleksander Madry, Aleksandar Makelov, Ludwig Schmidt, Dimitris Tsipras, and Adrian Vladu. 2018.
Towards deep learning models resistant to adversarial attacks.

JED: this is the subsection we need to create. It should discuss all the automated tools before noting that one way to test stuff is a human armed with an explanation (which is where this paper focuses)\\
Be sure to insert a citation to mutant agent paper and mention expl resolution, the SE folks will like that

JED@RKD: write about Groce

RKD@JED: Added in the.txt file 
}

\subsection{Human-AI Relationships in AI Systems}

\boldify{What are the relationships between people and AI?  These researchers found relationships x,y,z.}

When people are assessing an AI or engaged in some other AI-based task, what are their relationships with the AI?
Papachristos et al.~\cite{papachristos2021people} found four categories of how humans perceive and adopt their role in human-AI interactions; the two most pertinent to \emph{assessing} an AI are the ``Guide'' case and the ``Assistant'' case.
In their Guide case, the human takes a back-seat, intervening only when AI confidence scores slipped.
In their Assistant case, the human adopts a strictly dominant role to the AI, which assists with classification only when human confidence in their own abilities wane.
As an example of a domain where people preferred the dominant role (Assistant), consider DuetDraw \cite{oh2018lead}, a human-AI interactive drawing application providing the two described roles.
Those authors report that the preference for Assistant role was stronger when the application provided ``detailed'' explanations.

Another human role between the extremes of Assistant and Guide is that of teammate, in which the human and AI system collaborate with each entity playing to their own strengths in (hopefully) complementary ways (e.g.,~\cite{bansal2019updates,cooke2021effective,liang2019implicit}).
When faced with the prototype of an AutoAI system, data scientists suggested a collaborative relationship with the AI system instead of the fully automated design~\cite{wang2019human}.
This role attempts to mimic collaboration in human teams.
Molenaar~\cite{molenaar2022towards} introduces an example of this role in their ``six levels of automation model'' for personalized learning.
In this role, the teacher ``monitors'' and the technology oversees specific tasks, but the technology can still defer to the teacher and cede control.
The customer service domain has also utilized this type of human-AI relationship, or ``hybrid intelligence,'' so that AI and human customer service agents augment each other~\cite{wiethofais}.
According to Wang et al.~\cite{wang2021designing}, designing AI systems as \textit{tools} like other complex technologies, instead of encouraging collaborative relationships between human and AI systems, could be more beneficial.
A possible reason is that, in such collaborative environments, people judge their abilities and feel less competent than they actually are (i.e., their perceived effectiveness is less than the actual)~\cite{jacobsen2020perceived}.
In a similar case, data scientists had more confidence in the model they created by themselves in notebooks than the more accurate model they generated with AutoDS, an AI system~\cite{wang2021autods}.
Jacobsen et al.~\cite{jacobsen2020perceived} posits that this mismatch between perception and reality would persist if people are not able to get instant assessment of the collaborative decision.

\boldify{These are the guidelines to help design these relationships (**probably about 3 parags, one each for amershi, shneiderman, and wang/lim.}

To foster better collaborative human-AI relationships, AI systems designs should communicate clearly, provide explanations, and offer control of their actions to users~\cite{amershi2019guidelines, anderson-diversity-2024, shneiderman2020human}.
Copying existing human behavior and relationships is inadequate~\cite{shneiderman2020humanfresh}, it is important to leverage unique human and computer features and allow for different control levels based on the specific tasks~\cite{shneiderman2020human}.
For example, the scheduling system designed by Cranshaw et al.~\cite{cranshaw2017calendar} uses a 3-tiered technique for assigning tasks, assigning to the system tasks like creation of meetings; while the human takes care of updating meeting information or rescheduling events.
Researchers have found that providing information about the system's data source and decision process allows for easier adoption, productive use, and may help calibrate expectations---whether the stakes are high (e.g., detecting prostate cancer \cite{cai2019hello}) or low (e.g., scheduling events \cite{kocielnik2019will}).

\FIXME{MMB@anyone: Above parag (and anywhere else in Background that has this issue), needs to clarify which things are findings/results/facts, vs which things are the authors' opinions.  eg, in the 1st sentence, it says "AI systems designs SHOULD" -- make clear if that's a fact, or just what researchers believe?  (What I know: For Anderson it's a fact (finding), for Shneiderman I don't know but I'm guessing it's an opinion, and for Amershi2019 it's sort of in-between -- but if you add on the li2023assessingGuidelines reference it probably becomes a fact (you'd have to check which individual guidelines got confirmed in that paper)}


\subsection{AI Explanations}

\boldify{define AI explanation and mention its possible goals}

Examples of AI models that are particularly difficult to explain because of their complexity include neural networks (NN), ensemble models, and support vector machines (SVM).
These AI models feature in a lot of critical systems, so there are earnest efforts to make them more explainable while also maintaining high accuracy.

\boldify{Explanations have different types of info; eg, Lim et al's intelligibility types.}

Explanations can provide different types of information, which Lim et al.~\cite{lim2009why} formalized into a set of ``intelligibility types.''
Intelligibility types have proven helpful as XAI researchers design and evaluate their explanations' content.
For example, Lim et al. found that people preferred ``Why'' information for unexpected behavior, but in other contexts people have prioritized ``What'' information such as in strategy games~\cite{lim2009why, penney2018toward} and smart homes~\cite{castelli2017happened}.
The Why and Why Not intelligibility types have attracted the most attention from XAI researchers, examining explanations targeted at these types in a variety of domains, including pervasive computing~\cite{vermeulen2010pervasivecrystal}, email classification~\cite{kulesza2011oriented}, database queries~\cite{bhowmick2013not,he2014answering}, news feeds~\cite{cotter2017explaining}, and robotics~\cite{hayes2017improving,lomas2012robots,Rosenthal2016robots}.

Explanations for how an AI model generates output can be presented in different formats (e.g., text~\cite{yessenalina2010automatically,liu2019towards, hendricks2021generating}, saliency-maps~\cite{simonyan2013deep, selvaraju2017grad, brahimi2018deep}, graphs~\cite{kulesza2011my, kulesza2015principles}, and timelines~\cite{van2004explainable, core2006building}, etc.).
Based on the ``focus'' of the explanation in a single system, they can either be local (providing information about a particular decision) or global (providing information about how the whole system works)~\cite{mueller2019explanation, mueller2021principles}.
Also, there are three types of techniques used for explanations: opaque box, transparent box, and hybrid.

\subsubsection{Opaque Box Explanations}

Opaque box explanation techniques treat the whole AI model as opaque and operate at the input/output level of various decision instances.
Earlier research efforts such as LIME~\cite{ribeiro2016should}, SHAP~\cite{lundberg2017unified},  and LORE~\cite{guidotti2019factual} focused on generating local explanations.
LIME~\cite{ribeiro2016should} perturbs the input at a decision and examines how the output changes, sometimes fitting a simpler model to the decision boundary the observer views because of the perturbations.
SHAP~\cite{lundberg2017unified} generates Shapley values using different approximation methods that tell how important all features are at a particular decision.
LORE~\cite{guidotti2019factual} generates a decision tree classifier for input instances in the area around an AI system's particular decision.
From this decision tree, the system records two observations:
a rule that represents the path from the input to the decision output; and a set of rules that represent input changes that would cause a different decision.
Setzu et al.~\cite{setzu2021glocalx} extend LORE to a global context.
They created a ``model-agnostic'' explanation solution called GLocalX that hierarchically merges similar rule-based local explanations to build a global explanation of how the whole system works.
This ``global explainer'' is a transparent and simpler version (and possible replacement) of the opaque model with comparable accuracy.
Another opaque box explanation technique that provides global explanations is BEEF~\cite{grover2019beef}, which solves a combinatorial optimization problem to obtain ``balanced'' explanations.
Hendricks et al.~\cite{hendricks2016generating} introduced a user-centric opaque box technique for image classification.
The proposed AI model is made up of a standard deep classifier and an explanation mechanism that generates the explanation sentences using LSTM and a loss function that favors inclusion of class-specific information.
The output result is a ``visual explanation'' that contains attributes of the predicted class as well as the image description.
The main advantage of opaque box approaches is that they do not require access to the inner workings of the system---either because it is too complicated or because the user lacks access permissions (e.g., GPT-3~\cite{dale2021gpt} is only accessible through API calls).

The main drawbacks of these approaches are twofold.
First, each AI decision a user inspects requires the explanation system to re-run.
Additionally, fitting a second (ostensibly simpler) model to the original model's input-output pairs does not \emph{introspect} on the original model.
Thus, while opaque box approaches are useful for understanding the model, they are ill-suited for debugging.

\subsubsection{Transparent Box Explanations}

Transparent box explanation techniques reveal the internal structures and operations of the AI model.
An example of this technique that works for image classification is the deconvolutional network (deconvnet) method introduced by Zeiler and Fergus~\cite{zeiler2014visualizing}.
The deconvnet model \emph{``maps the feature activity in intermediate layers to the input pixel space''}~\cite{zeiler2014visualizing}.
When an image input is initially passed to a CNN during training, each layer generates features to pass as an input to the deconvnet model in that layer.
This method outputs a visualization that shows the parts of the input image that are informative for classification.
Similarly, in Deep Q networks, the learned features  ``map'' to different regions \cite{zahavy2016graying}.
Upon examination of these regions, applicable rules and policies are identifiable, forming an explanation of the AI model.
A different approach for CNNs, called network dissection~\cite{bau2017network, bau2020understanding}, also helps measure the degree of its ``interpretability'' by: 
(1) collecting a list of human-categorized concepts
(2) computing the features activated in each hidden unit for each concept
(3) aligning concept-activation pairs.

Aside from the feature activity in each layer, one could visualize granular level information such as each neuron's ``multiple facets''~\cite{nguyen2016multifaceted}.
Weidele et al.~\cite{weidele2019deepling} modified this type of visualization system with interaction components such that users could explore ``What If?'' scenarios.

The main advantage of transparent box explanations is their ability to assess and debug the model.
The main drawbacks relate mostly to scale, leading Sarkar~\cite{sarkar2022} to ask \textit{``Is explainable AI a race against model complexity?''}

\FIXME{MMB@anyone: In parag above, "their ability" -- needs to be clarified WHOSE ability -- the explanations'? }

\subsubsection{Hybrid Explanations}
\label{secBackgroundHybrid}

Hybrid explanation techniques combine and/or blend characteristics of opaque and transparent box approaches.
The hybrid explanation mechanisms possess a combination of characteristics (to a degree) from the opaque box and transparent box explanation techniques.
They do not necessarily operate exclusively on the AI model's inputs/outputs and may not be concerned about the AI model's granular internal structures, like neurons.
An example is  De et al's.~\cite{de2020explainable} ``Cluster-TREPAN.''
This approach first clusters the outputs at the hidden layer level.
Then, it applies TREPAN~\cite{craven1996extracting} to each cluster such that a decision tree that represents each cluster and produces both a set of rules and reason codes.
Once an AI's decision instance belongs to a hidden layer cluster, if the cluster-level decision tree's class prediction matches the neural network's, then the decision tree rule \emph{is} the explanation.

We utilized a hybrid approach in our work, using a NN and explanations drawn from prior work~\cite{dodge2022people}.
While we do not inspect individual neurons or layers, the whole model is designed to be more inherently explainable than mapping inputs directly to outputs.
To that end, the NN we adopted from prior work represents a function of the form in Equation~\ref{eqnScoresForall}, as opposed to Equations~\ref{eqnOneAction} or \ref{eqnScores}:
\begin{eqnarray}
STATE &\rightarrow& ACTION\label{eqnOneAction}\\
(STATE,\; ACTION) &\rightarrow& SCORE\label{eqnScores}\\
STATE &\rightarrow& SCORES\;\; \forall\; ACTIONS\label{eqnScoresForall}
\end{eqnarray}
This means that a whole host of questions are answerable without needing to rerun the network, because the single forward pass generates data to use in a variety of explanations (e.g., why did it select option $A$ and not option $B$?).
The main drawback of this strategy is that this function is harder to learn.

\subsection{How XAI Researchers Evaluate AI Explanations}

\boldify{Researchers have evaluated explanations both analytically and empirically.
Analytical is...**This parag D2+**}

Researchers have evaluated explanations both analytically and empirically.
For the context of XAI, Hoffman et al.~\cite{hoffman2018metrics} terms these two evaluation approaches as evaluating ``goodness'' and ``satisfaction,'' respectively.
In the analytical/goodness approach, XAI \emph{researchers} are the main actors, comparing an explanation's content/structure/presentation against accepted criteria and established guidelines.
For example, an XAI researcher could consider where their explanation system follows Amershi et al.'s~\cite{amershi2019guidelines} human-AI interaction guidelines and where it does not.

\boldify{...and Empirical is...**This parag D2+**}

In contrast, explanation \emph{consumers} are the main actors performing the empirical/satisfaction approach, by seeing and using various instances of some type(s) of explanations as they complete a task~\cite{mueller2019explanation}.
Many XAI evaluations are empirical.
One example is Dodge et al.'s lab study, in which participants assessed an AI system's effectiveness using instances of several different types of explanations~\cite{dodge2022people}.
Those authors used the results of the participants' assessments to gain insights into each explanation type's strengths and weaknesses.
Another empirical example is Bondi et al.'s investigation into the effects of communication contents and style on the people in human-AI collaboration when performing an image classification prediction task~\cite{bondi2022role}.
Their results showed that participants performed better and provided more accurate results when the communication content was strictly that the AI model ``deferred'' to them.
Kulesza et al.~\cite{kulesza2015principles} investigated people working with explanations that instantiated several explanation principles, and used the results to evaluate the principles as well as the  explanations.
Khanna et al.~\cite{khanna2022quant} investigated people assessing an AI with the help of explanations with/without a scaffolding approach called After-Action Review for AI (AAR/AI), and found that the AAR/AI-scaffolded explanations were significantly more effective than the same explanations without the scaffolding.
Other examples of the empirical approach abound; Lai et al.~\cite{lai2021towards} survey over 100 more.

\boldify{Explanations help users build mental models, and Prediction is a reasonable measure of mental models -- and of the expls that users built them from.**this parag is D2**}

\FIXME{JED@RKD: add citation to Eiband here, consider reprhasing the DO.

RKD@JED: Added in the .txt file}

Designing an approach for evaluating a particular AI explanation depends on what the explanation is trying to help a user \textit{do}.
....\cite{eibandFIXME}
Explanations have different purposes, but most have some connection to helping users build mental models of how the AI works.
A mental model is an in-the-head representation that an individual generates based on their prior experiences~\cite{johnson1989mentalModels}.
Users ``execute'' (in their heads) these models to understand and explain system behaviors, and predict the system's future behaviors, even for inputs/situations they have not yet seen~\cite{mueller2019explanation}.
Thus, evaluating how well users can \emph{predict} an AI's behaviors is one way XAI researchers can measure the quality of a user's mental model.
In XAI, mental models stem from explanations, so measuring the quality of a user's mental model is also a measure of the efficacy of a particular type of explanation to that user.
As we discuss next in this paper, we base our measurements on participants' prediction activities.

\section{Data Collection Methods}
\label{sectionMethods}

\subsection{Study 1 - MNK Games Domain}

\boldify{We recruited N participants at OSU via email.
Here is how we applied inclusion/exclusion, scheduled, and structured the study}

We recruited 86 participants at Oregon State University to complete  IRB-approved tasks in-lab via a flier distributed over email and posted on campus.
The only inclusion criteria were that participants be 18 or older and not study Computer Science, since we were interested in the perceptions of AI non-experts.
Once eligible participants gave informed consent, we scheduled them for a two-hour session, then randomly assigned them to one of eight treatments.

\boldify{treatments and tasks are defined as follows}

\begin{figure*}
    \centering
    \includegraphics[width = .8\linewidth]{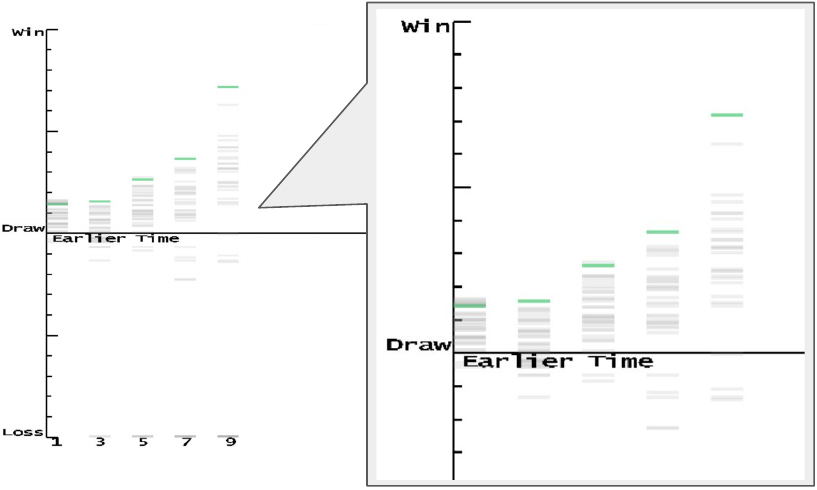}

    \vspace{2mm}\hrule\vspace{2mm}
    
    \includegraphics[width = .8\linewidth]{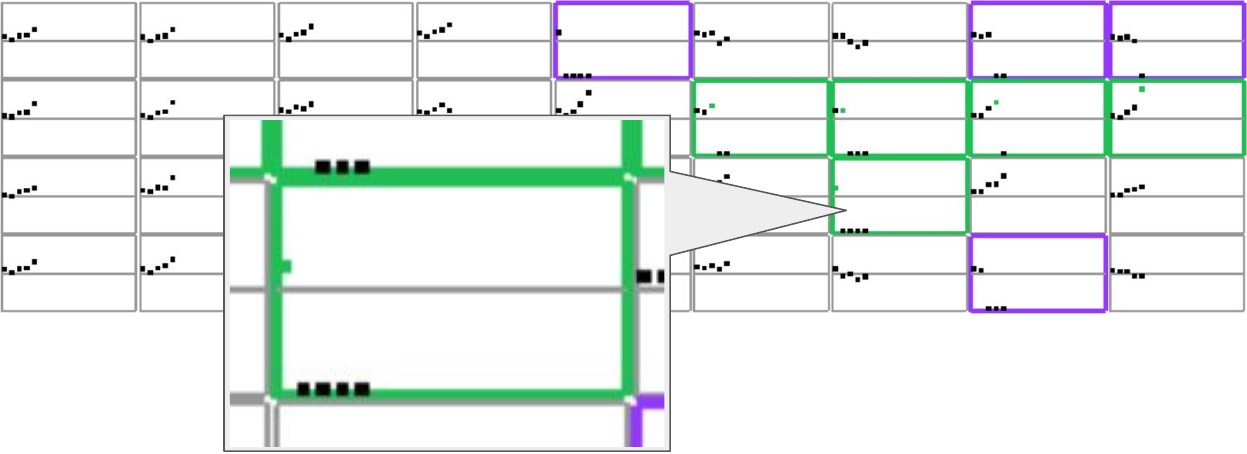}

    \vspace{2mm}\hrule\vspace{2mm}
    
    \includegraphics[width = .8\linewidth]{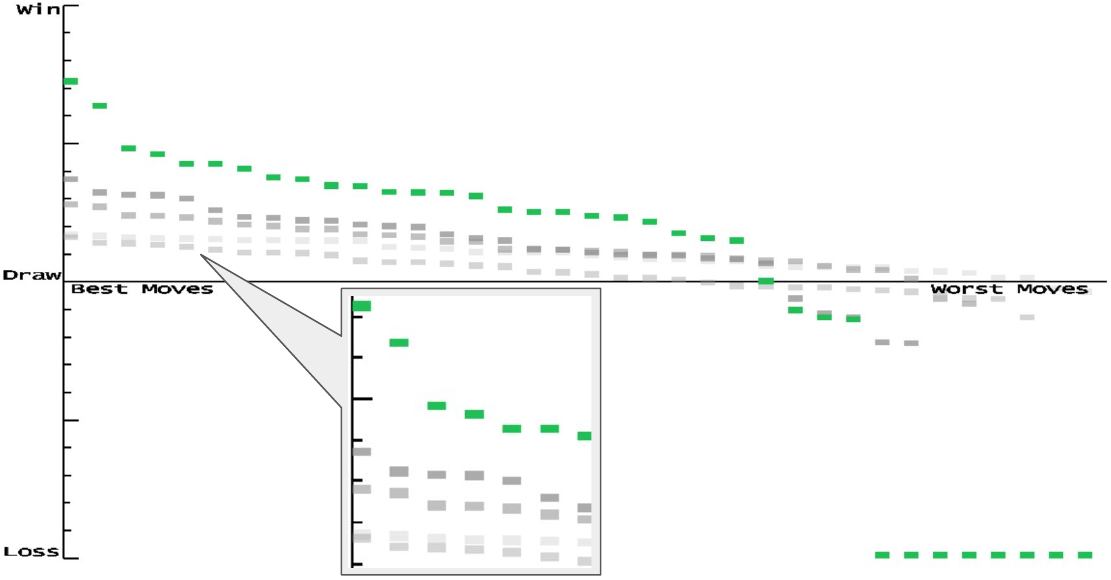}
    \caption{Explanations adapted from Dodge et al.~\cite{dodge2022people}.
    \textbf{Top:} \STTlong{} (\STT{}); \textbf{Middle:} \OTBlong{} (\OTB{}); and \textbf{Bottom:} \BTWlong{} (\BTW{}).
    Insets are \emph{NOT} part of the interface, but we provide them for greater figure clarity.
    }
    \label{figureExpls}
\end{figure*}

The treatments are based on combinations of the three explanations from Dodge et al.~\cite{dodge2022people} (shown in Figure~\ref{figureExpls}).
Treatments could have no explanations (\NONE{}, our control group), one explanation (\STT{}, \OTB{}, \BTW{}), two explanations (\STT{}+\OTB{}, \OTB{}+\BTW{}, \STT{}+\BTW{}), or all three explanations (\ALL{}).
This study employed multiple tasks\footnote{We also randomly assigned the order of the Ranking, Comparison, and Prediction tasks to each participant.
Regardless of the task ordering, each participant saw the same games and outcomes within a particular task.}, but this paper will focus on the Prediction Task, as described by Muramatsu et al.~\cite{muramatsu2001transparent}.

\subsubsection{Domain}

\boldify{we used MNK games, which has the following rules. Specifically, 9-4-4. Reasons are...}

Our domain was MNK games, which are a generalization of the well-known board game Tic-Tac-Toe (3-3-3).
In MNK games, each player alternates placing their piece ($X$ or $O$) in an effort to arrange their pieces in a sequence of length $K$ on a board with a size of $M \times N$.
In our study, we used 9-4-4, in which players attempt to create a 4 length sequence on a $9 \times 4$ board.
Thus, the action space in this study was 9x4x4=144 possible actions---that is, each time a participant predicted the agent's next move, they had to choose 1 out of 144 possibilities.

The MNK domain offers the following advantages for us in human lab studies:
1) the move tree has a limited depth because the board will ultimately fill, 
2) it offers robust empirical controls, and
3) despite having a comparable representation to domains like Go, its rules and strategy are significantly easier to understand, which makes tasks both accessible to laypersons and short enough for lab studies.

We implemented the game using an adaptation of Dodge et al.~\cite{dodge2022people}'s source code, which offers a simulator for the straightforward transition model of MNK games.
This simulator encodes each board using three states per square---controlled by opponent, controlled by the agent, and empty---in just two bits.

\subsubsection{Agent}
\label{subsec:Procedure}

\boldify{We used the same explanations as in prior work, as well as their agent, which has an architecture like this...
}
This section, which essentially follows Dodge et al.'s~\cite{dodge2022people} methodology, provides a brief overview of those authors' three interactive explanations (Figure~\ref{figureExpls}).  
The agent has-a convolutional neural network (CNN) to construct these explanations by predicting outcome tuples $O = (Win\%,\; Loss\%,\; Draw\%)$ for each square, given the $M \times N$ board.
The network features a two-channel input layer with the opponent's pieces always in channel two and the agent's pieces always in channel one, resulting in a tensor with dimension $M \times N \times 2$ (never visible in the interface).
The agent is CNN-based, with 8 total layers, producing an output of shape ($M \times N \times O$)~\cite{dodge2022people}.
To select actions, the agent begins by performing a forward pass on the network, then flattening the outcome dimension via a generalized value function  (proposed by Sutton et al.~\cite{sutton2011}, though still used, e.g.,~\cite{lin2021contrastive}).

\subsubsection{Explanation 1: \STTlong{} (\STT{})}
\label{subsubsec:StTime}

\boldify{The first explanation is STT, it works as follows}

The \STT{} explanation focuses on the temporal aspect of the data in an effort to address the question, 
\emph{``How did the agent score each square at each decision?''}.
Time is the X-axis in this explanation. Every time the explaining agent makes a decision, a new column appears for the user, showing the agent's evaluation of \emph{each} square at that choice.
Consider Figure~\ref{figureExpls}(top), where each column of the explanation shows the {\color{green}Green} agent's evaluation of each of the 36-square board's squares. 
For each decision, one rectangle in each column has the same color as the agent (in this case, {\color{green}Green}), representing the score for the square that the agent chose.
When the user hovers over a square on the gameboard, they see that square's score highlighted in every column of the \STT{}; 
similarly, hovering over the scores on the \STT{} highlights the square associated.
If the user highlights scores that overlap, the interface highlights every square associated with \emph{every} highlighted score.

\subsubsection{Explanation 2: \OTBlong{} (\OTB{})}
\label{subsubsec:OnBoard}

\boldify{The second explanation is OTB, it works as follows}

The \OTB{} explanation emphasizes a connection between the temporal and spatial dimensions of the data, attempting to answer: \emph{``How effective is this square at various times?''}.
As Figure~\ref{figureExpls}(middle) shows, each square of the \OTB{} explanation is itself a mini \STT{}, showing the scores for that square through time.
When the user hovers a square on the gameboard,  a single \OTB{} chart with the same coordinates is highlighted;
similarly, hovering a chart on the \OTB{} highlights a single square on the gameboard.
Any time the explaining agent makes a decision, each of the 36 charts receives one new data point with the score for that square.

\subsubsection{Explanation 3: \BTWlong{} (\BTW{})}
\label{subsubsec:BtoW}

\boldify{The third explanation is BTW, it works as follows}

The \BTW{} explanation places emphasis on the data's value dimension, in an effort to address the question, \emph{``How did the agent score its choices at each timestep?''}.
Storing all the values for all squares at all times results in a tensor with both space and time dimensions.
This explanation cuts along the time dimension, sorting all 36 squares by descending value.
Figure~\ref{figureExpls}(bottom) shows how each decision results in a single data series, with only the most recent being colored in the agent's color ({\color{green}Green}) and interactive (hovering a score highlights a square and vice versa).
The grey colors given to the old data series get increasingly light to illustrate how the score distribution shifts with time.

\subsubsection{Procedure}

\boldify{Participants had 4 tasks and touched a lot of agents along the way, this is how we made agents and also made them discernible from each other}

During the study's three tasks, participants observed a total of 12 agents, created via mutant agent generation~\cite{dodge2022people}.
Every task contained 4 games with controlled randomization so each participant saw the same moves, and participants had 5 minutes for the first 3 games and 9.5 minutes for the last game.
To keep all the participants working at the same pace, we used lock dialogues\footnote{See supplemental materials}.
Once the allotted time expired, a researcher provided a 2-letter password to everyone in the room, allowing participants to proceed to the next game/task.
After each game, participants would fill out a paper form containing questions from AAR/AI~\cite{dodgej2021aarai}.
Specifically, the form would collect: \emph{``What happened in this game (write any good, bad, or interesting things you've observed in these past moves and/or games)? (a few sentences)''} and \emph{``Is there anything in the explanation that helps you understand \textbf{Why} the AI you're assessing did the things it did? Please specify which explanations you are referring to. ($\sim$2 sentences)''}.
We did not include the rest of the questions from AAR/AI to reduce participant fatigue.
Participants kept these forms for reference until the end of a task, at which point researchers collected the old forms and gave fresh ones for the next task.

\boldify{Participants were to predict and AI's actions (Predict), compare two AI and say which is better (Compare), and they were to rank 4 agents from the ``best'' to ``worst'' (Rank).
Here's how the tutorial went}

Participants first went through a researcher-led tutorial to better understand the MNK game and our tasks.
In the tutorial, participants played a game against the {\color{green}Green} agent.
Then, they would observe {\color{green}Green} playing {\color{purple}Purple}, and the researcher described each of the explanations and tasks.
Before starting the tasks, we told participants they would be observing games and using the information available on the interface to assess the agents.

\boldify{Alright reader, we know you know they're doing 3 tasks, but we're focused on 1, and here's how it went...}

\begin{figure}
    \centering
    \includegraphics[width = 0.6\linewidth]{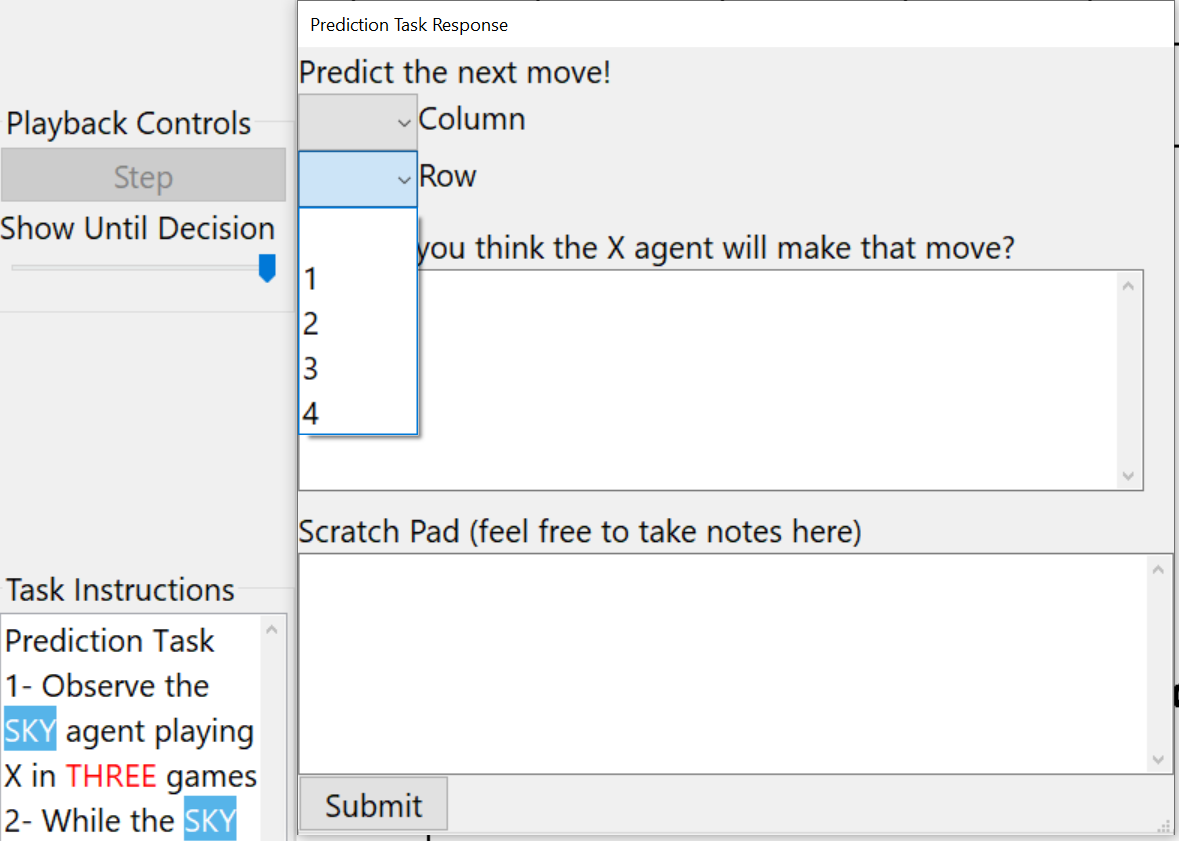}
    \caption{An example of how participants saw the dialogue box (right) and made their predictions.
    Notice that they selected coordinates of their prediction via drop-down menus, to specify a column letter and row number.
    Once they had formulated a prediction, they provided their justifications in the first text box, with an optional scratch pad below.
    While they were making their predictions, the ``Step'' button was not available (left, top), but they could Rewind via the slider (left).}
    \label{fig:prediction-dialogue-box}
    
\end{figure}

Although participants performed three tasks, as mentioned previously, the scope of this paper covers \textit{only} the Prediction task.
During this task, participants first watched the {\color{cyan}Sky} agent play three games against three yet-unseen opponents,  using the ``Step'' button to proceed through the games and the ``Rewind Slider'' to revisit previous states as desired.
Participants could also hover over squares or scores associated with them.
After both agents made their first move, we asked participants which square they thought the {\color{cyan}Sky} agent would take next and \emph{why} they thought it would take that move, using the dialogue box in Figure~\ref{fig:prediction-dialogue-box}.
Participants could see all information about the previous moves the {\color{cyan}Sky} agent made, including the explanation(s), but they could not see information about the \textit{next} move until they submitted their prediction. 

\boldify{And finally, the study was over, what did that look like}

Upon finishing all tasks, participants filled out a post-task questionnaire and received \$20 USD compensation.

\subsection{Study 2 - Analysis in Four Towers Domain
}
\label{sectionStudy2}

Our second analysis used data  produced by Anderson et al.'s earlier study~\cite{anderson2020mental, Anderson2019} in a size 4 action space---each time a participant made a prediction, they needed to chose 1 out of 4 possibilities.
These data allowed us to compare results from a very small action space against the larger 144-action space in Study 1.
This section summarizes Anderson et al.'s methodology~\cite{anderson2020mental, Anderson2019}.

\subsubsection{Domain}

The player controls a tank with a kite-like design that is positioned in the middle of a map with four quadrants.
The player's objective is to damage and destroy adversaries in order to score points.
However, if it attacks allies, it loses points. 
The player has four actions available, attacking one of four quadrants; whether it contains an adversary (black) or ally (white). 
The object's quadrants can contain: big/small forts (forts can return fire), cities/towns (the agent always loses points when attacking non-combatants), and enemy tanks.

\subsubsection{Agent}

The agent playing the game used reinforcement learning (RL) to learn a Q-function that calculates the anticipated total reward of executing a given action in a given state, then following the policy.
The policy was based on a decomposed Q-function~\cite{lin2021contrastive, sutton2011} to divide the reward into various components based on the reward type (e.g., damaging an enemy vs taking damage), with the goal of revealing more relevant information about the agent's preferences.
In order to learn the decomposed Q-function, the SARSA algorithm~\cite{Russell:2003:AIM:773294} incrementally changed each reward component according to the agent's experience.
The agent represented the Q-function with neural networks which accept the current game state as input and estimate the Q-value for a specific reward component.
After training, the agent chose actions that would maximize the total Q-value, which is the sum of the Q values for all reward components.

\subsubsection{Explanation 1: Reward Bars}

Anderson et al.~\cite{anderson2020mental, Anderson2019}'s first explanation presented the reward decomposition as a bar chart.
For a specific action, the chart contained a bar cluster where each bar shows the Q-value contribution of a particular reward component.

\subsubsection{Explanation 2: Saliency Maps}

Anderson et al.~\cite{anderson2020mental, Anderson2019}'s second explanation used saliency maps to show input features the agent concentrates on most when making decisions.
These maps showed the relative weight of several game state components using a heatmap overlay.
Those authors adopted a perturbation-based saliency maps created by altering game objects, such as enemy tanks, then evaluated how much the output Q-values were impacted, and finally visualized the result as a heatmap.

\subsubsection{Procedure}

Anderson et al.~\cite{anderson2020mental, Anderson2019}'s study contained four treatments: saliency maps, reward decomposition bars, both, or no explanations (control).
Each of the 124 participants took part in a 2-hour lab experiment, all of whom were not computer science majors.
Before each session, the researchers provided a tutorial describing the game mechanics, explanations, and interface.
After that, participants watched as the trained agent made 14 decisions.
Along the way, participants predicted which quadrant the agent would attack next and explained their reasoning in an open-ended manner at each step.
After submitting their prediction, participants viewed the agent's actual decision and the accompanying explanation. 

\subsection{Statistical Analysis}

In this paper we test hypotheses via:
ANOVA (when we have both equivariance and normality); and
Kruskal-Wallis (when we have equivariance but not normality).
We verify the equivariance assumption with Levene's Test and the normality assumption with the Shapiro-Wilk Test.

\section{Proposed Analysis Methods for Prediction Task Data}
\label{sectionAnalyzing}

\boldify{We divide our strategies into two main buckets; those yielding a distribution and those yielding a single score.}

\begin{figure}
    \centering
    \includegraphics[width = .4\linewidth]{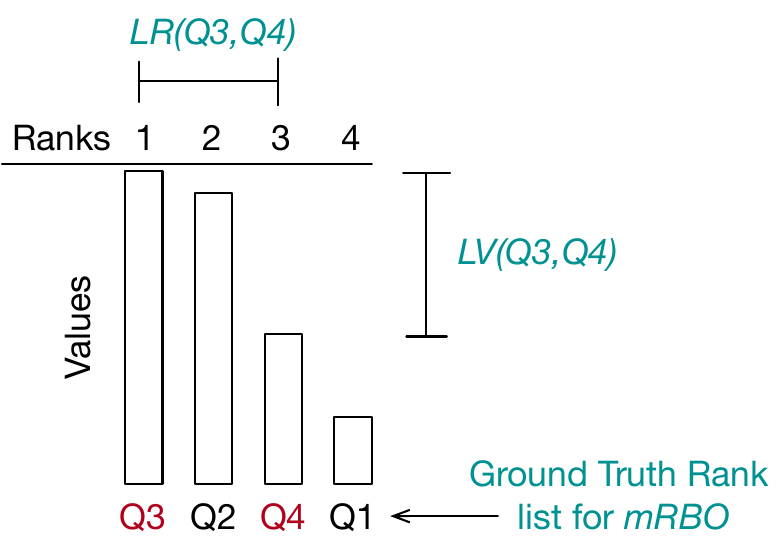}
    \caption{A notional illustration of our three measurement constructs in teal text.
    Suppose an agent in the Four Towers domain predicts values as the bar chart shows (sorted in decreasing order, as in the \emph{BTW} explanation shown in Figure~\ref{figureExpls}).
    The agent will select Q3 (shown in red) for this decision, since it has the max value.
    Suppose a particular participant predicted Q4 (also in red), $LV(Q3, Q4)$ would compute the difference in value space as shown.
    Now, in rank space, $LR(Q3, Q4) = 2$ as shown, because regardless of the numerical values predicted, this is their order in the action list sorted by value.
    $mRBO$ grabs exactly that rank list and compares it against some reference list.
    In our work, we combine predictions a group of participants in a voting schema to create a rank order, but the rank order could also come from asking each participant for a (partial) ordering of what they think the agent will prefer.
    }
    \label{figNotionalMath}
\end{figure}

This section will give our mechanistic answer to \emph{RQ1 - How can we operationalize analyzing prediction task data in a way that has some degree of independence from domain and model?}.
After we see the results of applying these approaches, we will return to RQ1 in Section~\ref{secDiscussion} to discuss the efficacy of our proposed mechanisms.
All our strategies are based on the notation presented in Table~\ref{tableMathNotation}, and we divide them into two categories; those yielding a distribution and those yielding a single score.
We have also provided a notional illustration of our measurement constructs in Figure~\ref{figNotionalMath}.

\begin{table}

\begin{tabular}{@{}l | l@{}}
\textbf{Variable} & \textbf{Description}
\\\hline
$P$ &
A set of participants
\\

$A$ &
The set of all actions
\\\hline

$\hat{a}$ &
The action that the agent selected (correct answer)
\\

$a_i$ &
The action predicted by the $i$th participant $\in P$
\\\hline

$V(a\in A)$ &
The value estimated by the agent for an action
\\

$R(a\in A)$ &
The rank of value estimated by the agent for an action 
\\\hline

$D(\cdot)$ &
A discretization function, such as the letter grade system we described in Footnote~\ref{footnoteGrade}
\\

\end{tabular}
\caption{Notation for the math presented in Section~\ref{sectionAnalyzing}}
\label{tableMathNotation}

\end{table}

\subsection{Strategies Yielding a Distribution}
\boldify{In which contexts are these desirable?}

The main advantage of a distribution is support for computing comparative statistics.

\subsubsection{\underline{L}oss in \underline{V}alue (LV)}
This strategy leverages the idea that two actions are similar if the agent assigns them similar values, meaning the agent perceives them to have similar \emph{outcomes}.
In our case, values are interpretable: in Study~1, values represent ``advantage,'' measured by $Win\%-Loss\%$; in Study~2, values represent the number of points the agent expects to obtain.
\begin{equation}
LV(\hat{a}, a_i) = V(\hat{a}) - V(a_i)\label{eqnLV}
\end{equation}

\subsubsection{\underline{L}oss in \underline{R}ank (LR)}
However, values in general need not possess any intuitive semantic equivalent; in such instances, it could be better to utilize the agent's values' rank ordering, since ranks are a simple and widely applicable concept.
This strategy leverages the idea that two actions are similar if the agent assigns them values which are similar in the rank ordering of \emph{all} values.
However, ranks have a disadvantage since they destroy exact relationships between the predicted values.
\begin{equation}
LR(\hat{a}, a_i) = R(\hat{a}) - R(a_i)\label{eqnLR}
\end{equation}

\subsubsection{\underline{D}iscretized \underline{L}oss in \underline{R}ank (DLR)}
\label{sectionDiscretize}

This strategy is based on Loss in Rank, but discards some detail by applying a discretization function, which can be helpful for interpretation and visualization.
Discretizing uses a customizable number of bins, so in our case to analyze Study~1, we defined $D(\cdot)$ with six bins based on deciles~\footnote{A decile is one of 10 groups resulting from division of according to some variable into 10 equal groups.
In our case, there were 36 moves, so predicting one of the agent's rank 1--4 moves would place the participant in the first decile, rank 5--8 in the second, and so on.
This grading system is just one possible strategy for operationalizing partial credit.
For example, participants who guessed in the top 10\% of all moves (i.e., ranks 1---4) received an ``A'' for that prediction.
Participants who guessed in the next 10\% (i.e., ranked 5---8) received a ``B'', and so on until a ``D'' grade for guessing between moves 13---16.
Participants received a ``F'' for their prediction if they guessed moves that the agent ranked worse than $17^{th}$.
\label{footnoteGrade}}
in a common grading system found in U.S. education (e.g., 90\%+ earned an A, 80--89\% a B, 70--79\% a C, 60--69\% a D, and $<60$\% an F).
Note that we did not apply Equation~\ref{eqnDLR} to Study~2 because the very small action space of size 4 means it already uses very few bins.
\begin{equation}
DLR(\hat{a}, a_i) = D(LR(\hat{a}, a_i)) = D\bigl(R(\hat{a}) - R(a_i)\bigr)\label{eqnDLR}
\end{equation}

\subsection{Strategies Yielding a Single Score}

The main advantage of a single score is that it is easy to understand and present.
Clearly, the distributions of $LV()$, $LR()$, and $DLR()$ will have measures of central tendency (e.g., Tables~\ref{tabMNKmonolithic} and \ref{tabFourTowersMonolithic} show the mean).
However, we have an additional single-score producing metric to propose, the $mRBO$.

\subsubsection{\underline{M}odified \underline{R}ank-\underline{B}iased \underline{O}verlap between agent's preferences and participants' group-wise preferences ($mRBO$)}
\label{sectionMRBO}

This strategy infers two preference rankings and compares them by computing a metric similar to Rank-Biased Overlap ($RBO$)~\cite{Webber2010rbo}.
The first rank list comes directly from the agent, defined by computing the $R(\cdot)$ function $\forall a \in A$.
The second comes from a group of participants, using the counts of each prediction within the group to impose an order on the actions.
Our approach to constructing the second ranking is essentially a voting schema, meaning it is likely that many squares will be tied at 0 votes, which we truncate from the ranking.

$RBO$ produces an output in the range $[0,1]$, where 0 means the rank lists are disjoint and 1 meaning they are identical.
$RBO$ is defined on: lists $S$ and $T$; in addition to hyperparameters  $p$  (which controls weighting for high ranking values), and $k$ (the evaluation depth) as follows (Equation 23 from~\cite{Webber2010rbo}):
\begin{equation}
RBO_{EXT}(S,\; T,\; p,\; k) = \frac{|S_{:k}\cap T_{:k}|}k\cdot p^k + \frac{1-p}p\sum^{k}_{d=1}\frac{|S_{:d}\cap T_{:d}|}{d}\cdot p^{d}
\label{eqnRBO}
\end{equation}

The reason that we modified the $RBO$ metric is because, after we applied it off-the-shelf, we observed it behaving in counter-intuitive ways, which we will detail later in Section~\ref{sectionRQ2}.
Our modifications\footnote{Webber et al.~\cite{Webber2010rbo} showed a lot of properties about the the $RBO$ equations described therein.
To us, the modifications described in this paper seem unlikely to have harmed any of their proofs, but we did not verify any of the properties those authors proved.} are fairly simple, shown in red in the next equation: 1) adjusting the denominator inside the sum to stop increasing when one of the lists runs out of elements; and 2) changing the denominator of the first term in the same way\footnote{See an implementation of our modified version in source provided within our Supplemental Material.}.
Without loss of generality, assume that $S$ is the shorter of the two lists.
Thus, our version is as follows, where $k = |T|$:
\begin{equation}
mRBO_{EXT}(S,\; T,\; p,\; k) = \frac{|S_{:k}\cap T_{:k}|}{\color{red}|S|}\cdot p^k + \frac{1-p}p\sum^{k}_{d=1}\frac{|S_{:d}\cap T_{:d}|}{\color{red}min(|S|, d)}\cdot p^{d}
\label{eqnMRBO}
\end{equation}

In our work we combined predictions from multiple participants to create a rank order, but there are other ways to create the reference list for $mRBO$.
One notable example could be requesting a longer preference ordering (whether partial or full) from a single participant, which would make the $mRBO$ approach yield a distribution.

\section{Results RQ2 - How well did participants predict in MNK games?}
\label{sectionRQ2}

Overall, our results suggest that \emph{binary} measurement of participants' prediction performance is too coarse-grained to answer this RQ convincingly.
However, the metrics proposed in this paper reveal a bit more, as we will see.

\subsection{Adopting the binary prediction framing}

\begin{figure}
    \centering
    \includegraphics[width = .48\linewidth]{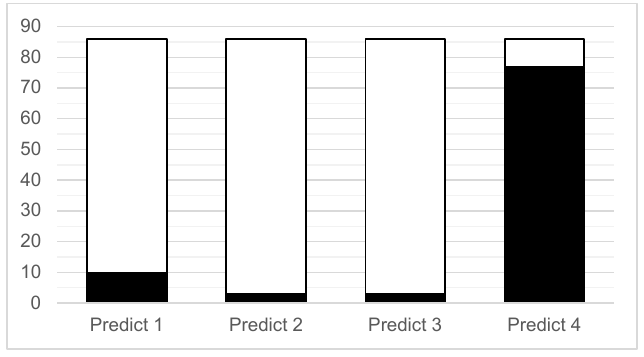}
    \hfill
    \includegraphics[width = .48\linewidth]{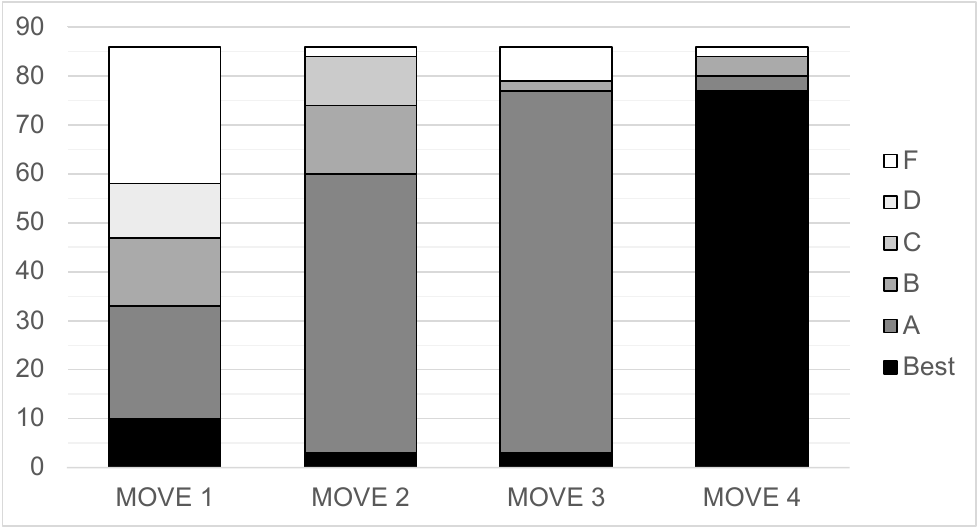}
    \caption{Illustration of the visual difference between the binary prediction framing vs one of the partial credit systems we propose applied to Study 1's data from the MNK games domain.
    \textbf{Left}: Overview of \emph{binary} prediction correctness for every participant, showing correct predictions in black and incorrect predictions in white.
    Both floor and ceiling effects are prevalent, which hinder comparative statistics.
    \textbf{Right}: Overview of distributions of grades $DLR()$ for every participant at every prediction.
    Unfortunately, floor and ceiling effects are still present (e.g., predictions 2 and 3 have moved from the floor to the ceiling).
    Only prediction 1 seems well conditioned, and that is where we will find our only statistically significant result.
    }
    \label{figMNKoverview}

    \FIXME{JED@SNK: I just noticed that these figures have different labels (predict vs MOVE). If possible, make consistent with the rest of the document, I don't think I have these figures}
\end{figure}

\begin{figure}
    \centering
    \includegraphics[width = .45\linewidth]{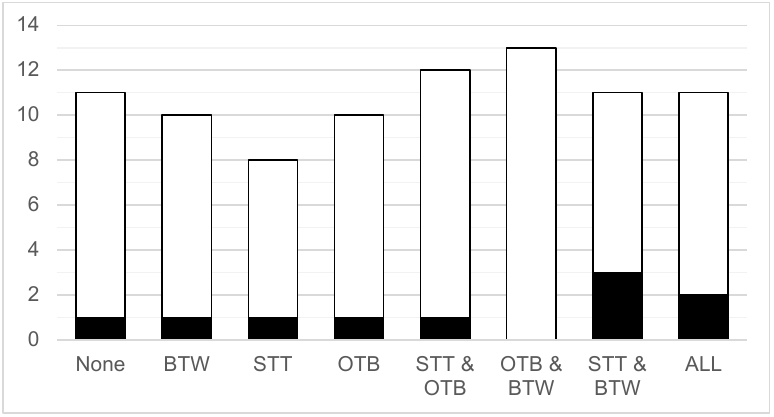}
    \includegraphics[width = .45\linewidth]{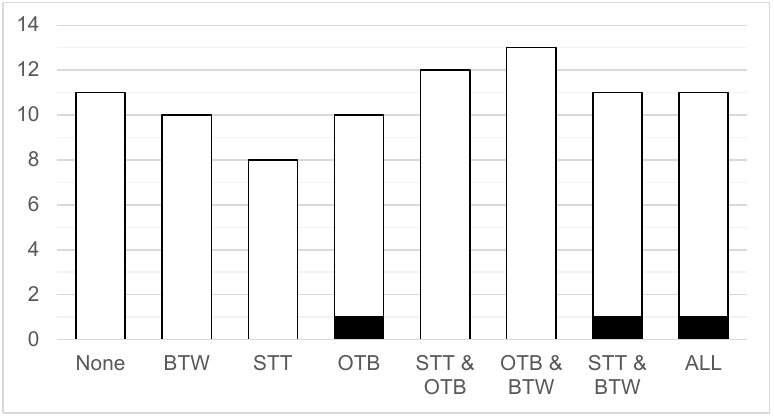}\\
    \includegraphics[width = .45\linewidth]{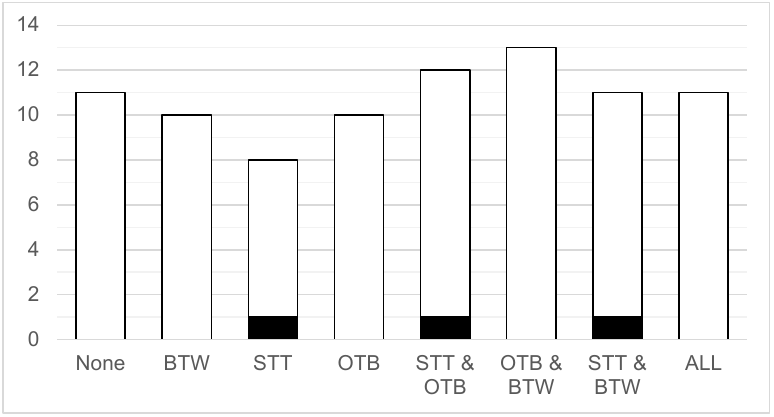}
    \includegraphics[width = .45\linewidth]{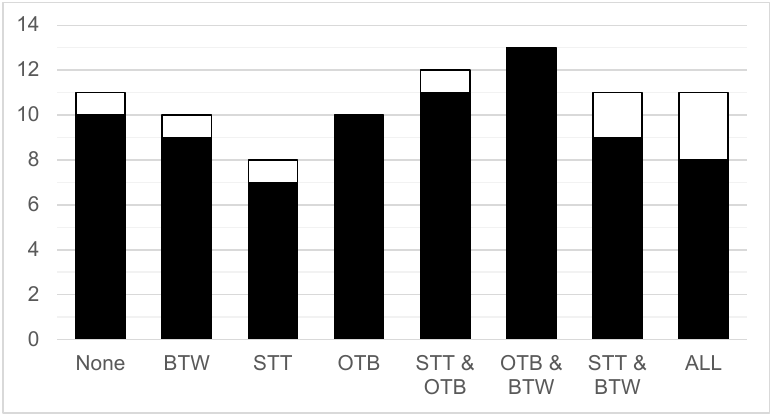}
    \caption{Every participant's correctness on every prediction, divided by treatment. 
    \textbf{Top left}: Prediction 1. 
    \textbf{Top right}: Prediction 2.
    \textbf{Bottom left}: Prediction 3. 
    \textbf{Bottom right}: Prediction 4.
    Now how the floor and ceiling effects we saw in the data overall are even worse when dividing by treatment.}
    \label{figureBinaryTreatment}
\end{figure}

It is troublesome but unsurprising that both floor effects and ceiling effects are prevalent in the data we collected. 
For example, Figure~\ref{figMNKoverview} shows that of the 86 participants in the experiment:
10 participants correctly made the first prediction (floor), 3 the second (floor), 3 the third (floor), and 80 the fourth (ceiling).
The same floor and ceiling effects are visible when we divide the participants into groups, as shown in Figure~\ref{figureBinaryTreatment}.
As a result of these floor and ceiling effects, we were not able to find any statisitically significant differences in these data.

The crux of the issue is that these measurements do not differentiate between incorrectly predicting what the agent ranked as the \textit{second-best} option, and predicting the \textit{thirteenth-best} option, as described in Section~\ref{sectionIntro}.

\subsection{Adopting the proposed framing, $LV()$ and $LR()$}
\label{secMNKlVlR}

However, when we switch over to using $LR()$ (Equation~\ref{eqnLR}), we observe some significant results in the data.
Applying this strategy to our data adds shades of gray to the black and white images shown in Figure~\ref{figMNKoverview} summarizing all participants' grades across the four predictions that they made.

\begin{figure}
    \centering
    \includegraphics[width = .49\linewidth]{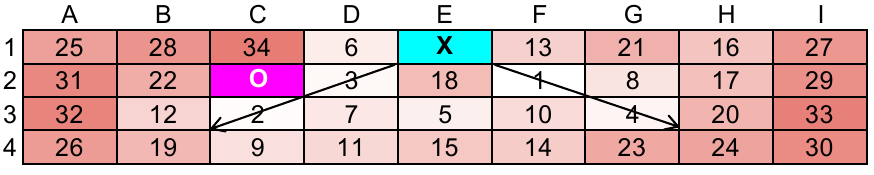}
    \hfill
    \includegraphics[width = .49\linewidth]{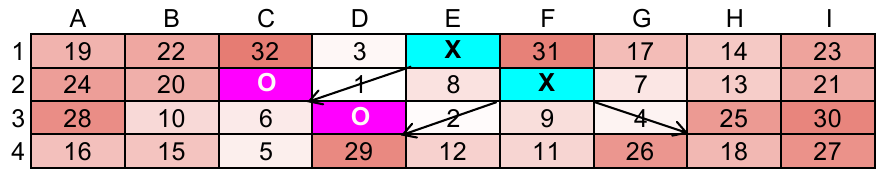}
    \\
    
    \includegraphics[width = .49\linewidth]{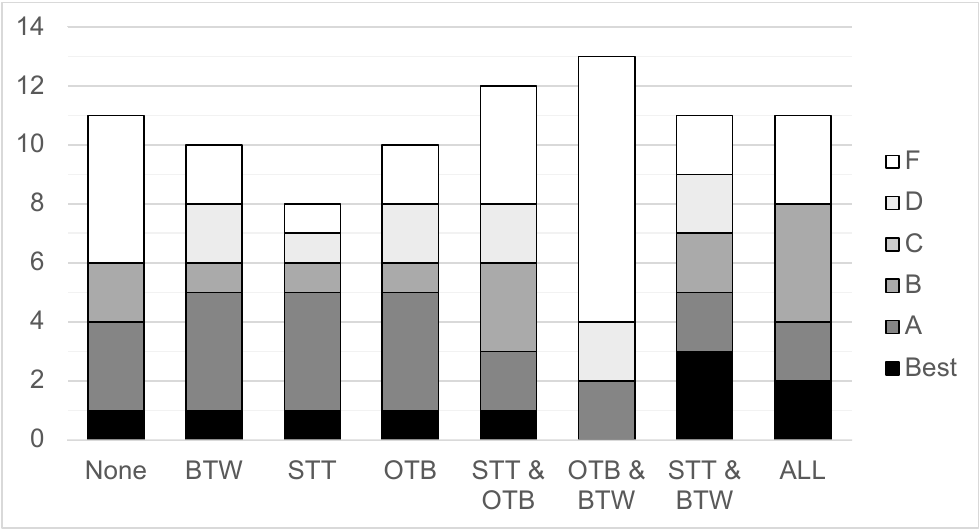}
    \hfill
    \includegraphics[width = .49\linewidth]{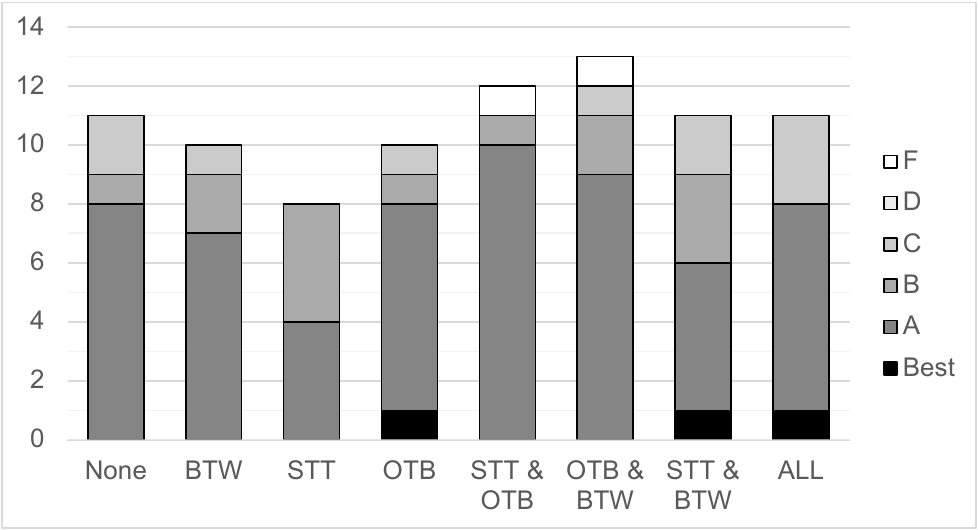}
    \\\vspace{10pt}
    \includegraphics[width = .49\linewidth]{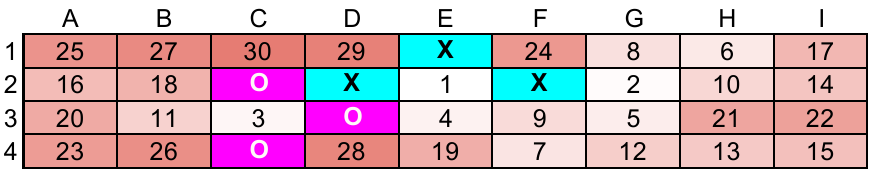}
    \hfill
    \includegraphics[width = .49\linewidth]{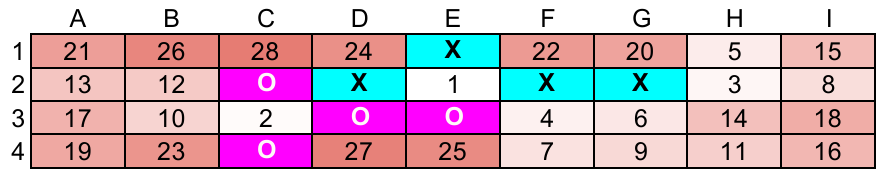}
    \\
    
    \includegraphics[width = .49\linewidth]{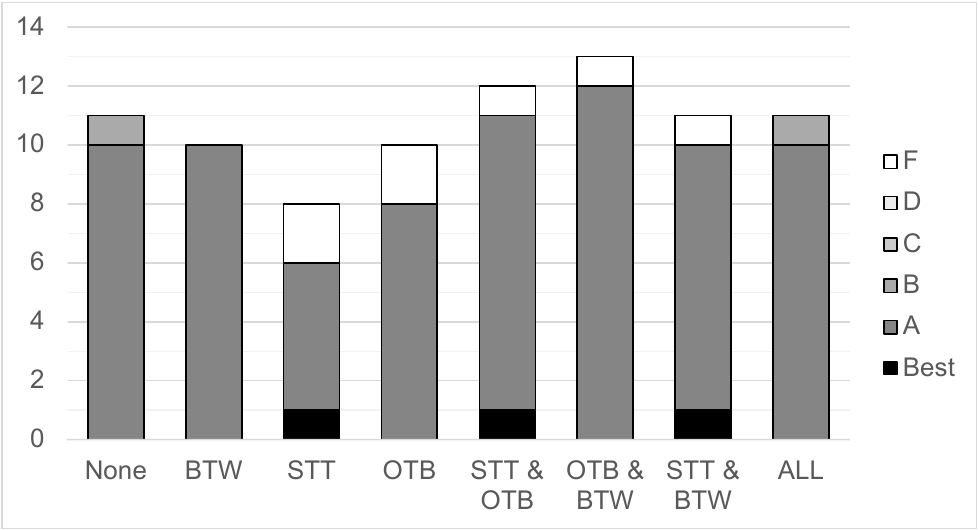}
    \hfill
    \includegraphics[width = .49\linewidth]{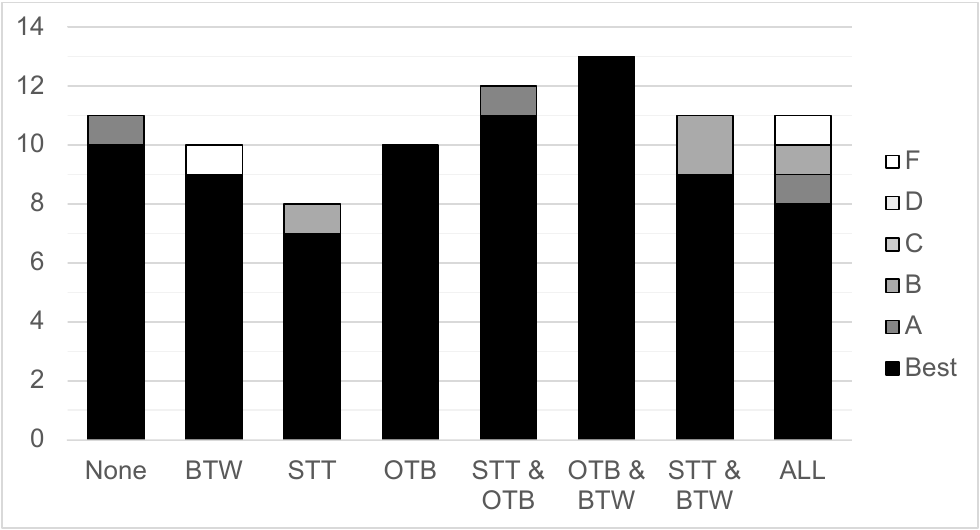}
    \caption{Agent ranks and Participants' graded predictions for each prediction.
    Comparing the bar charts with those found in Figure~\ref{figureBinaryTreatment} reveals the same overall prevalence of floor and ceiling effects.
    \textbf{(top left)} For the first prediction, the agent took F2, while valuing diagonal squares, as shown by the arrows.
    In this part of the figure, it seems like \OTB{}+\BTW{} is different from the others from visual inspection, which was not apparent in Figure~\ref{figureBinaryTreatment}.
    \textbf{(top right)} For the second prediction, the agent took D2, but 61.6\% of participants chose G3, which earned them an 'A.'
    \textbf{(bottom left)} For the third prediction, the agent took G2, though 44.2\% of participants selected E2.
    \textbf{(bottom right)} For the fourth prediction, 89.5\% of participants chose the best answer, E2, resulting in a clear ceiling effect.
    }
    \label{figGraded}
\end{figure}

As an example, in Figure~\ref{figureBinaryTreatment}, we saw that participants in the \OTB{}+\BTW{} treatment were the only ones in the first prediction where \textit{nobody} guessed the correct square.
However, it is difficult to infer much from that since participants in the other treatment groups did not do much better, with around \emph{one} person providing the correct prediction.
Now, consider that same \OTB{}+\BTW{} treatment's bar in Figure~\ref{figGraded}, which shows a lot of participants receiving F scores.

\begin{figure}
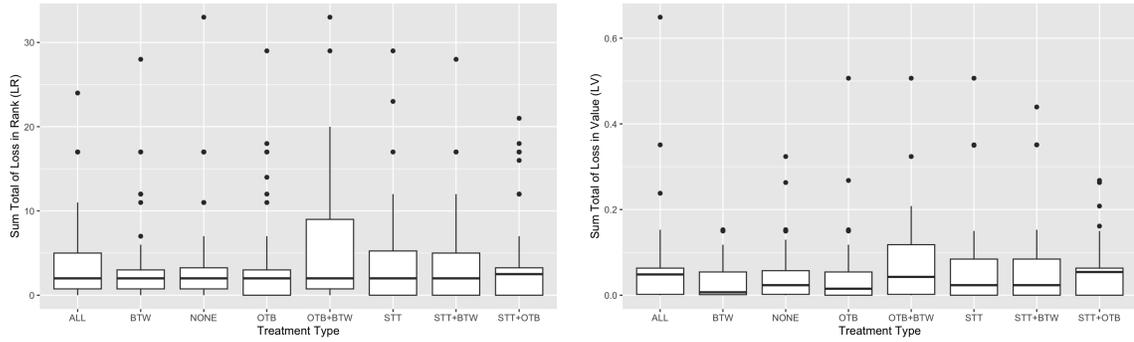

    \centering
    \includegraphics[width = .49\linewidth]{assets/MMLR.pdf}
    \hfill
    \includegraphics[width = .49\linewidth]{assets/MMLVOutlierRemoved.pdf}
    \caption{\textbf{Left}: Distributions of the \textit{loss in rank} (y-axis) based on all predictions from participants in each treatment (x-axis).
    \textbf{Right}: Distributions of the \textit{loss in value} (y-axis) based on predictions from participants in each treatment (x-axis).
    The higher the participants rank/value loss, the worse they performed, meaning up is bad on both charts.
    Similar to Figure~\ref{figureBinaryTreatment} (top-left), it is clear that the \OTB{}+\BTW{} treatment are guessing worse moves; in fact, \textit{significantly} worse moves.}
    \label{figP1ANOVA}
\end{figure}

Indeed, as the boxplots in Figure~\ref{figP1ANOVA} illustrate, participants in that treatment were predicting significantly worse squares, as compared to the agent's ranks (ANOVA, F(7,78) = 2.2084, p = .0423)\footnote{We know that these were the participants responsible for the significance, since removing them also removed the significance \newline(ANOVA, F(6,66) = 0.6613, p = .681).
Both of these statistical tests used the distribution from $LR()$ (Equation~\ref{eqnLR}), applied to the first prediction.}.


To understand the difference between the $LV()$ and $LR()$ computation, consider choosing the second best square at prediction~1 vs prediction~2.
Prediction~1's estimated best square\footnote{See our supplemental materials for the rest of the agent's value estimates, agent's value ranks are visible in Figure~\ref{figGraded}(top).}
has value 0.2118 and the second best 0.2009, while prediction~2's estimated best square has value 0.2438 and the second best 0.2011.
This means that in both cases the $LR()$ returns 1 (since both predictions correspond to the second-best square).
However, the $LV()$ returns 0.0109 for prediction 1 and the $LV$ for prediction 2 is \emph{four times higher} at 0.0427.
Since we can interpret value differences as a change in advantage (or $win\% - loss\%$),  the agent perceives prediction 1 to be a much closer decision.
As a result, we might expect the distribution for human predictions of the agent's behavior to reflect that.

\subsection{Adopting the proposed framing, $mRBO$}

\begin{figure}
    \centering
    \includegraphics[width = \linewidth]{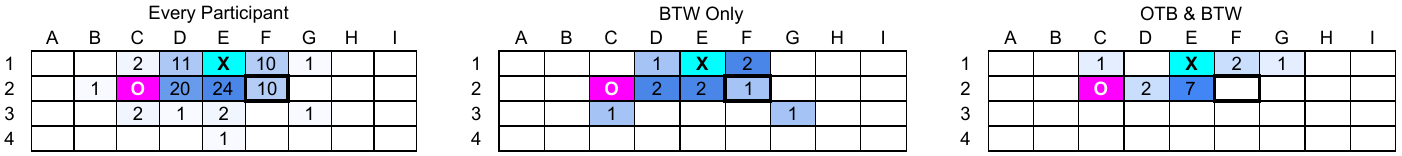}
    \caption{Heat map of predictions offered by participants in a few different treatments for the agent's first decision, where the agent took F2. 
    Darker blues means more participants predicted that square.
    Note the difference in distribution of choices across the board.
    \textbf{Left:} Every participant. 
    \textbf{Center:} \BTW{} only. 
    \textbf{Right:} \OTB{}+\BTW{}.}
    \label{figVotes}
\end{figure}

As described in Section~\ref{sectionMRBO}, $mRBO$ relies on having two lists to compare.
The agent's preference list is visible in Figure~\ref{figGraded}.
To create the other list, we select a group of participants to combine predictions in a voting schema.
One result of this grouping is visible in Figure~\ref{figVotes}), the rest are available in our supplemtal documents.
From the figure, it is clear that there are marked differences between the vote distributions of the treatment groups.

The result of constructing these voting groups and computing $mRBO$ to compare treatment groups is visible in Table~\ref{tabMNKmonolithic}.
The table includes the average $LV()$ and $LR()$ as additional reference points.

\begin{table*}
    \centering
    \begin{tabular}{@{}l | rrrrr | rrrrr | rrrr@{}}
    
    & \multicolumn{5}{c|}{mean $LV$}
    & \multicolumn{5}{c|}{mean $LR$}
    & \multicolumn{4}{c}{$mRBO$}
    \\
    \textbf{Treatment}
    & All & P1 & P2 & P3 & P4
    & All & P1 & P2 & P3 & P4
    & P1 & P2 & P3 & P4
    \\\hline\hline

    \NONE{}
    & .056
    & .100
    & .078
    & .021
    & .024

    & 4.4
    & 10.6
    & 4.8
    & 1.8
    & 0.2

    & .465
    & .507
    & .910
    & .955
    \\\hline

    \OTB{}
    & \textbf{.053}
    & .066
    & \textbf{.065}
    & .080
    & \textbf{0}

    & 4.2
    & 7.2
    & \textbf{3.9}
    & 5.7
    & \textbf{0}

    & .480
    & .635
    & .613
    & \textbf{1.000}
    \\\hline

    \BTW{}
    & .077
    & .064
    & .073
    & \textbf{.004}
    & .168

    & \textbf{3.9}
    & 7.1
    & 4.5
    & \textbf{1.3}
    & 2.8

    & .573
    & .513
    & \textbf{.955}
    & .550
    \\\hline

    \STT{}
    & .071
    & \textbf{.047}
    & .082
    & .111
    & .044

    & 4.6
    & \textbf{5.1}
    & 4.9
    & 7.6
    & 0.6

    & .527
    & .580
    & .511
    & .845
    \\\hline\hline

    \STT{}+\BTW
    & .065
    & .058
    & .081
    & .055
    & .064

    & 4.2
    & 6.5
    & 5.2
    & 4.1
    & 0.9

    & \textbf{.706}
    & .564
    & .689
    & .845
    \\\hline

    \STT{}+\OTB
    & .054
    & .084
    & .071
    & .040
    & .022

    & 4.3
    & 9.6
    & 4.3
    & 3.0
    & 0.2

    & .477
    & .557
    & .809
    & .955
    \\\hline

    \OTB{}+\BTW
    & .067
    & .136
    & .081
    & .052
    & \textbf{0}

    & 5.9
    & 15.4
    & 4.7
    & 3.7
    & \textbf{0}

    & .297
    & .545
    & .758
    & \textbf{1.000}
    \\\hline
    
    \ALL{}
    & .070
    & .065
    & .075
    & .027
    & .113

    & 4.1
    & 6.8
    & 4.7
    & 2.0
    & 2.7

    & .581
    & \textbf{.652}
    & .810
    & .776
    \\\hline
    
    \end{tabular}
    \caption{MNK Games domain results from applying Equations~\ref{eqnLV}, \ref{eqnLR}, and \ref{eqnMRBO} to each treatment group at each prediction and also combining all predictions.
    The best score in each column is bold, which shows that a single-explanation treatment outperformed all combination treatments except for three exceptions: a tie on P4, and $mRBO$ for P1 and P2.}
    \label{tabMNKmonolithic}
\end{table*}

Now for the reasons behind our modifications to $RBO$ (Equation~\ref{eqnRBO}).
As Table~\ref{tabMNKmonolithic} indicates, every participant in the \OTB{} treatment was correct for prediction 4, but the resulting participant rank list achieves an unexpectedly poor $RBO$ score of 0.57---despite the fact that this group of votes was perfect!
Similarly, the participants in \NONE{} treatment gave the most votes to the best square for prediction 4, while the third best square was the only other vote recipient.
For this prediction, the \NONE{} treatment received an $RBO$ score of 0.65---better than \OTB{} despite the votes being objectively worse.
Tables~\ref{tabMNKmonolithic} and \ref{tabFourTowersMonolithic} indicate that the $mRBO$ seems to show better match with intuition.

\section{Results RQ3 - How well did participants predict  in the Four Towers domain?}\label{sectionRQ3}

\boldify{So here is the ``right answer'' if you had full ex post facto access on the decision process}

In the course of our reanalysis of the data Anderson et al.~\cite{Anderson2019, anderson2020mental}, we performed a series of statistical tests to evaluate the underlying distributional properties and homogeneity of variances across four distinct treatments. 
Table~\ref{table:DecisionPoints} shows the ``right answer'' for each decision, as well as predicted values and vote counts.

\begin{table}
\centering

\begin{tabular}{@{}l|c@{}c@{}c@{}c@{}}
\multirow{2}{*}{\rotatebox{90}{\fontsize{9pt}{10pt} \selectfont Task 1~~~~~}}
                        & {\fontsize{8pt}{12pt} \selectfont DP1} 
                        & {\fontsize{8pt}{12pt} \selectfont DP2}  
                        & {\fontsize{8pt}{12pt} \selectfont DP3}  
                        & {\fontsize{8pt}{12pt} \selectfont DP4}  \\ 
                        & \includegraphics[width=3.64cm]{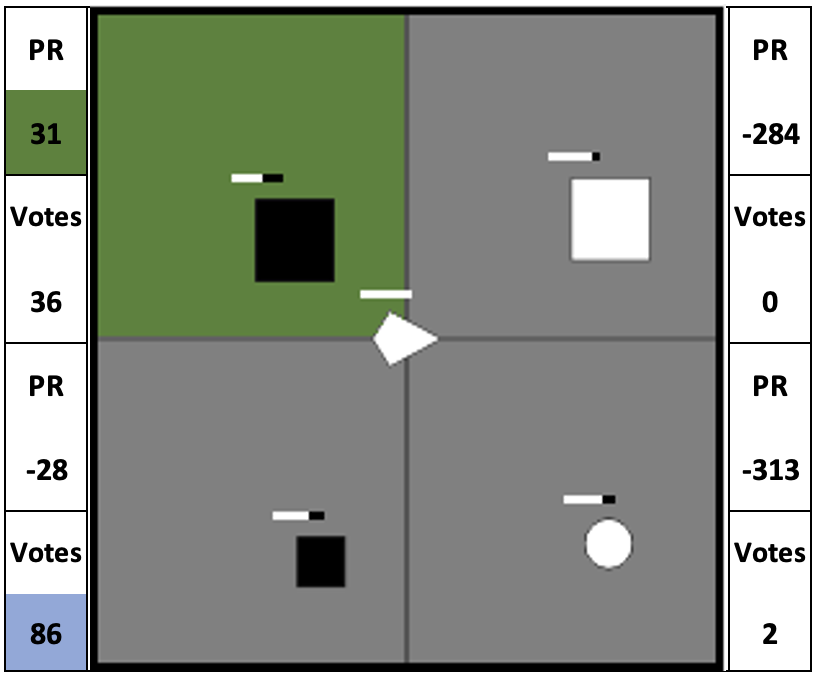}
                        &  \includegraphics[width=3.64cm]{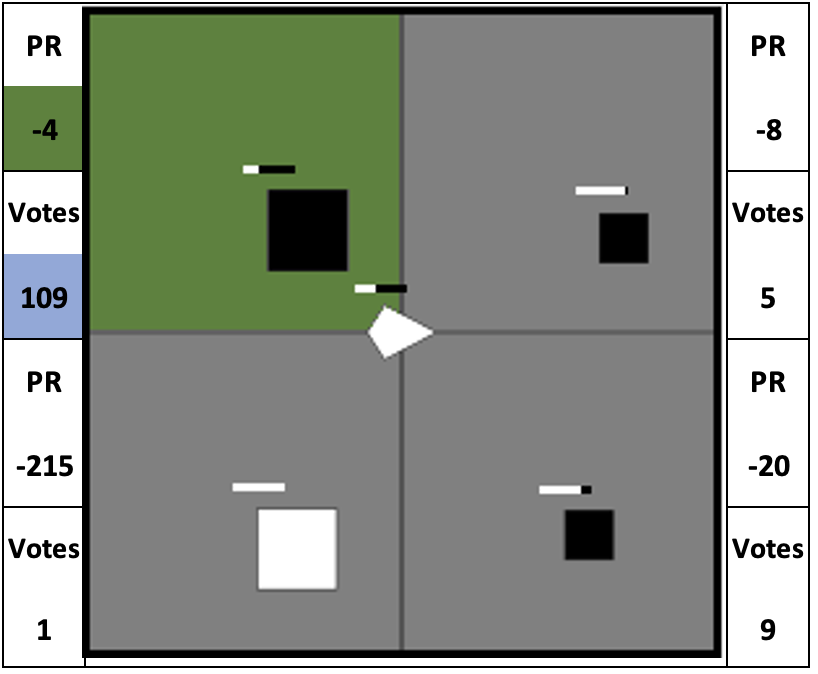}    
                        &  \includegraphics[width=3.64cm]{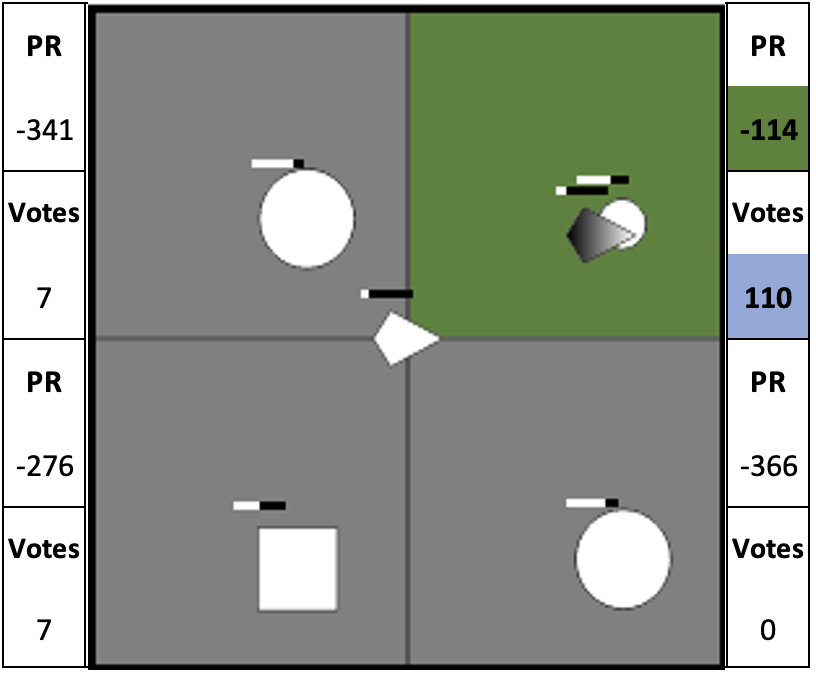}    
                        &  \includegraphics[width=3.64cm]{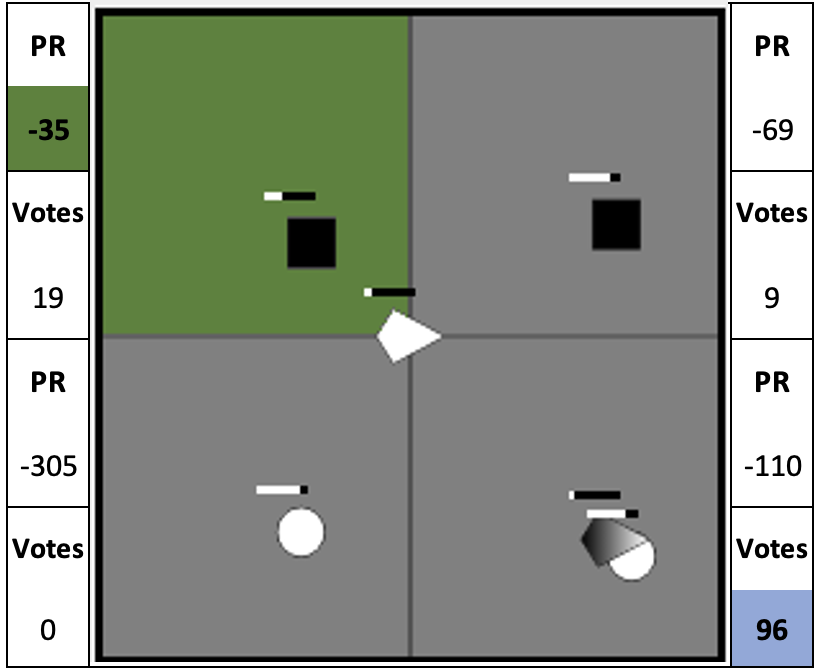}\\ \hline
\multirow{2}{*}{\rotatebox{90}{\fontsize{9pt}{10pt} \selectfont Task 2~~~~~}} 
                   
                        & {\fontsize{8pt}{12pt} \selectfont DP5}  
                        & {\fontsize{8pt}{12pt} \selectfont DP6}  
                        & {\fontsize{8pt}{12pt} \selectfont DP7}  
                        & {\fontsize{8pt}{12pt} \selectfont DP8} \\ 
                        &  \includegraphics[width=3.64cm]{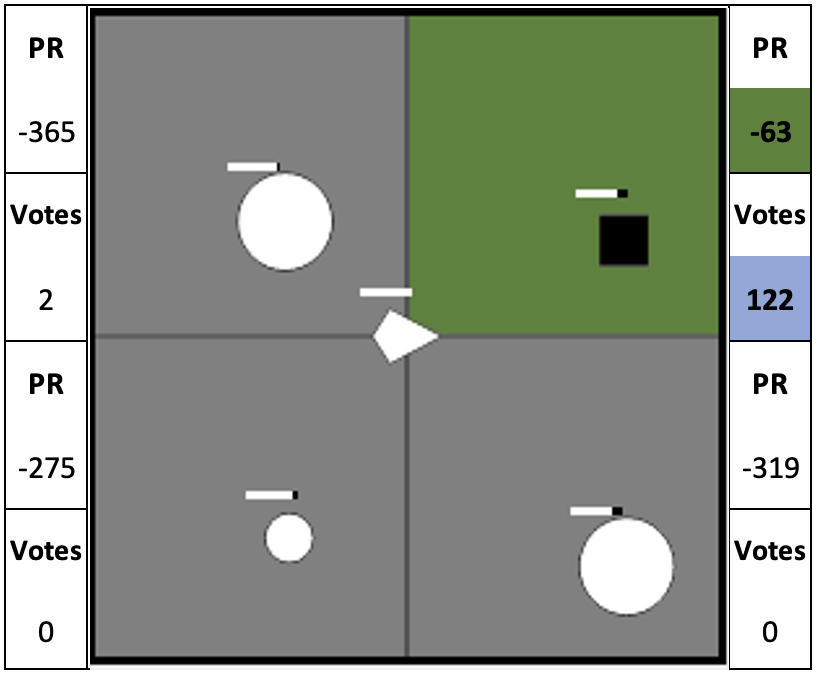}    
                        &   \includegraphics[width=3.64cm]{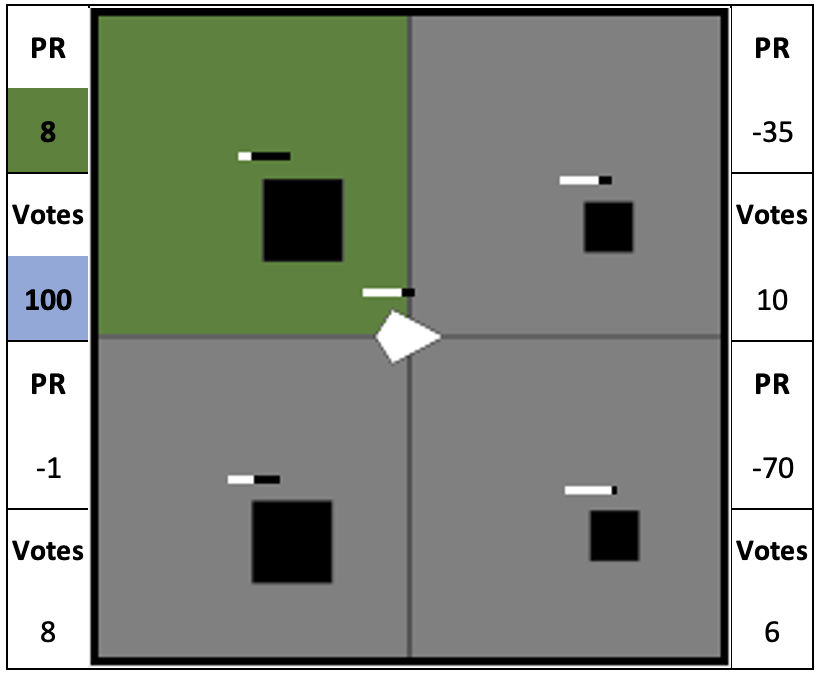}   
                        &  \includegraphics[width=3.64cm]{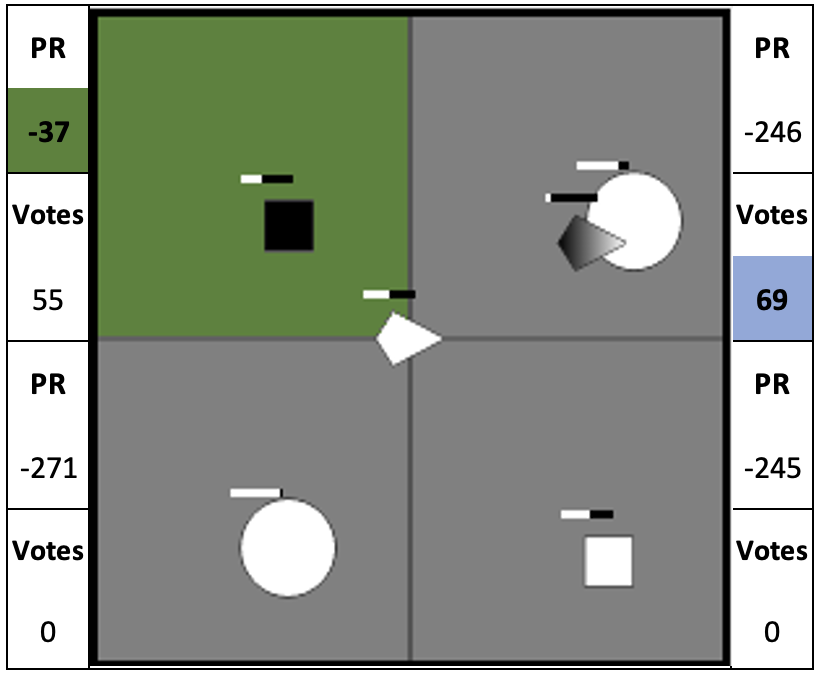}    
                        &   \includegraphics[width=3.64cm]{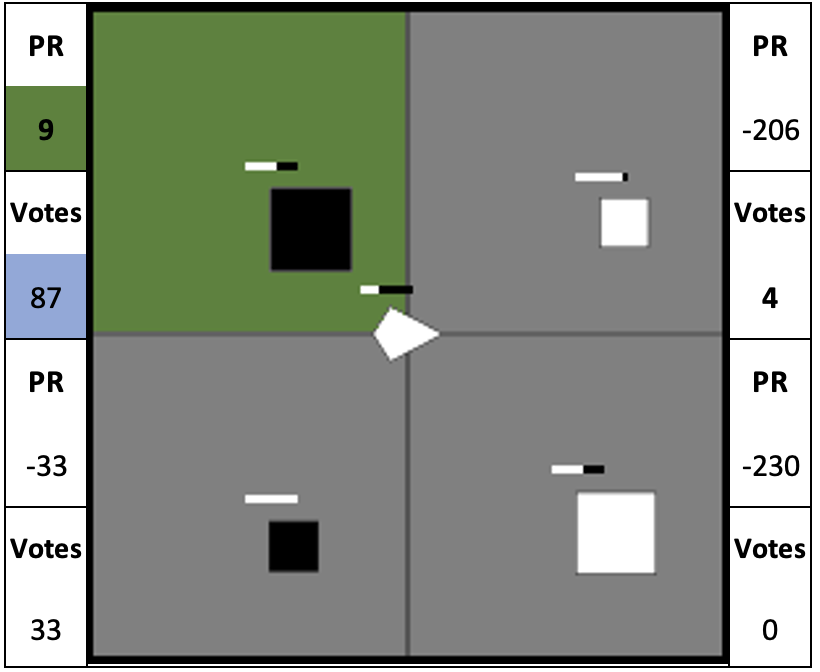}  \\ \hline
\multirow{2}{*}{\rotatebox{90}{\fontsize{9pt}{10pt} \selectfont Task 3~~~~~}} 
                        & {\fontsize{8pt}{12pt} \selectfont DP9}  
                        & {\fontsize{8pt}{12pt} \selectfont DP10} 
                        & {\fontsize{8pt}{12pt} \selectfont DP11} 
                        &  \\ 
                        &  \includegraphics[width=3.64cm]{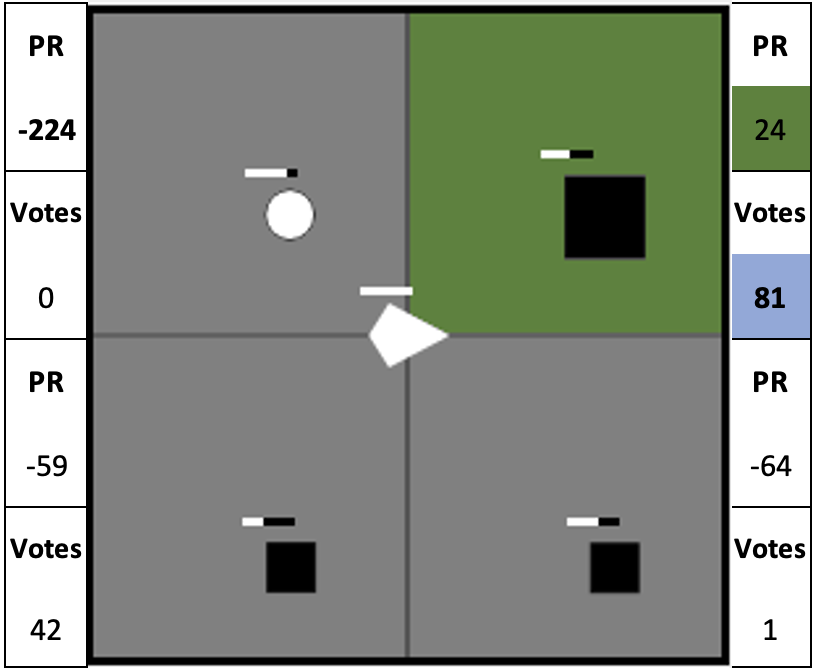}    
                        &   \includegraphics[width=3.64cm]{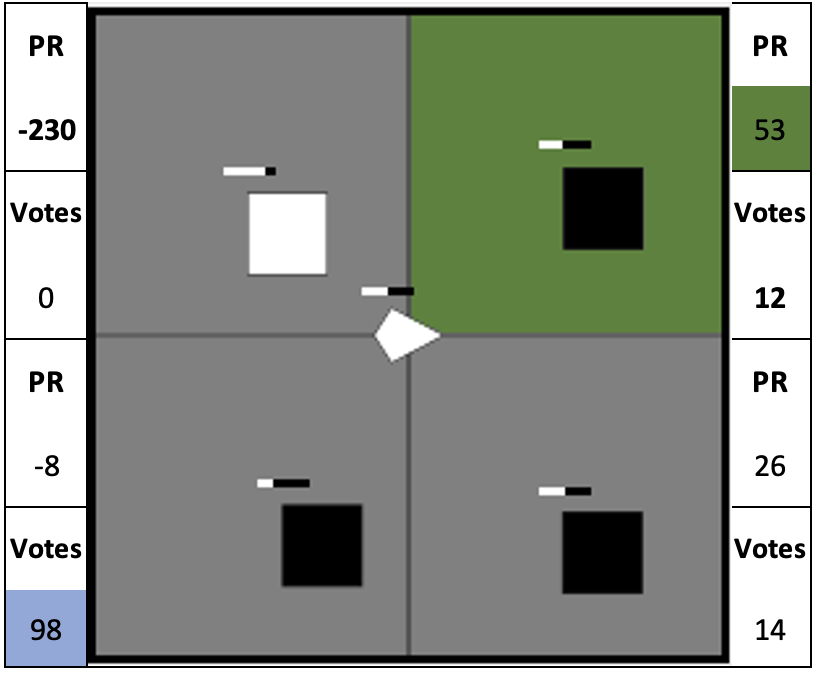}   
                        &   \includegraphics[width=3.64cm]{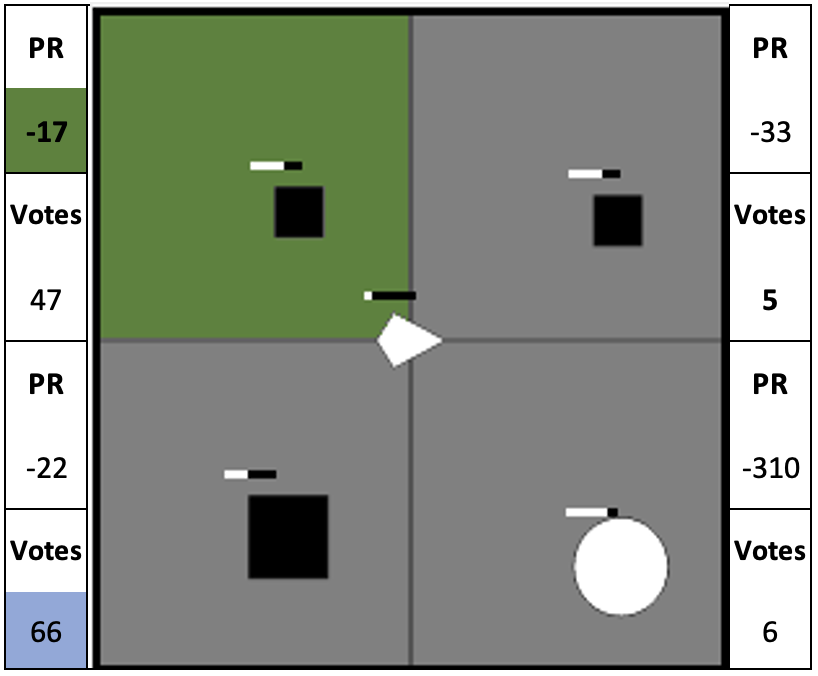}   
                        &    \\ \cline{1-4}
\multirow{2}{*}{\rotatebox{90}{\fontsize{9pt}{10pt} \selectfont Task 4~~~~~}} 
                        & {\fontsize{8pt}{12pt} \selectfont DP12} 
                        & {\fontsize{8pt}{12pt} \selectfont DP13} 
                        & {\fontsize{8pt}{12pt} \selectfont DP14} 
                        &     \\ 
                        &  \includegraphics[width=3.64cm]{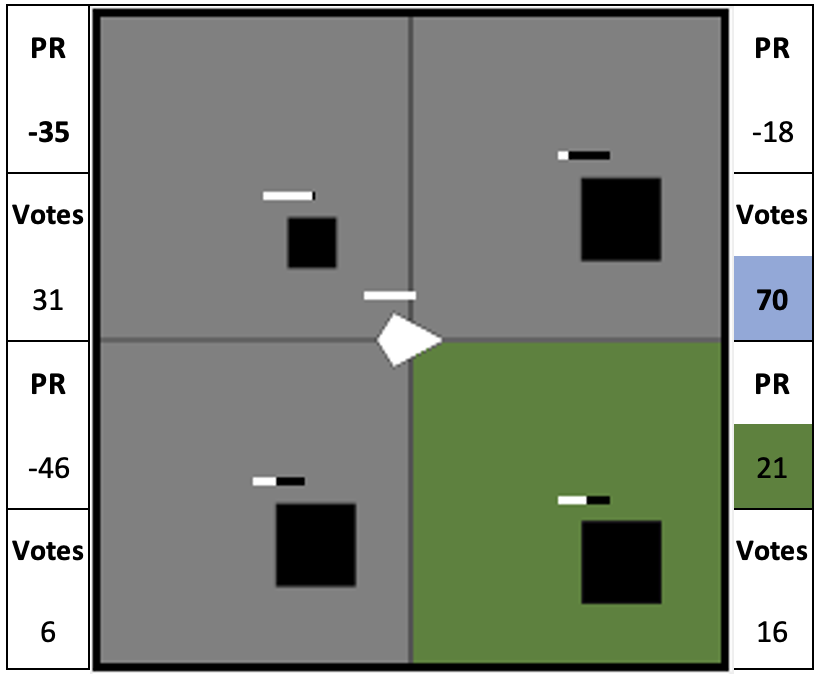}    
                        &   \includegraphics[width=3.64cm]{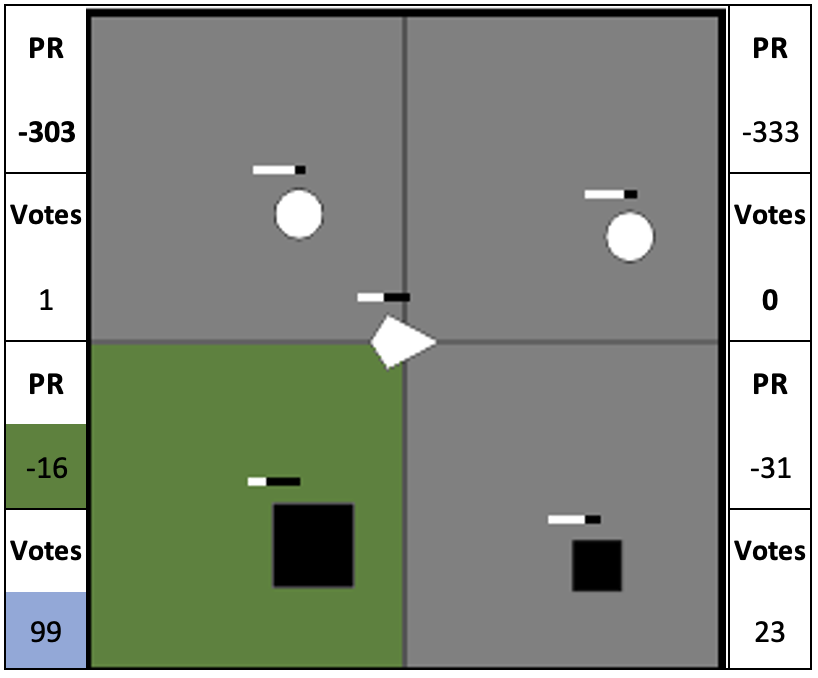}   
                        &   \includegraphics[width=3.64cm]{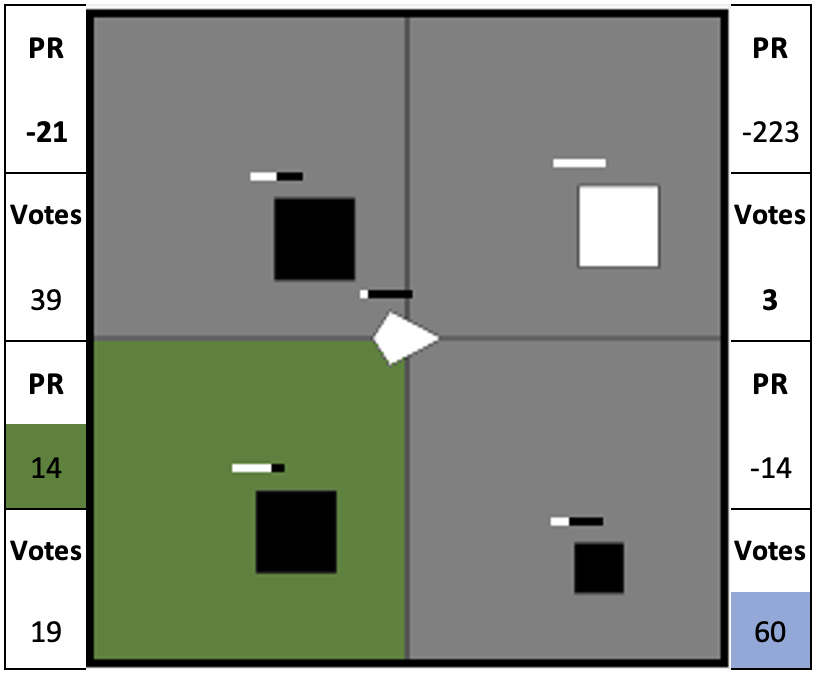}   
                        &   
\end{tabular}

    \caption{The tasks and their Decision Points (DPs). 
    We have highlighted the action the AI chose in green.
    At the edge of the map we provide the agent's predicted reward (PR), as well as the total number of votes participants in all treatments gave to that prediction.
    We have highlighted the most-predicted quadrant in blue, which does not always coincide with the green highlights.
    }
\label{table:DecisionPoints}

\end{table}

\subsection{Adopting the binary prediction framing}

\begin{figure}
    \centering
    \includegraphics[width = .4\linewidth]{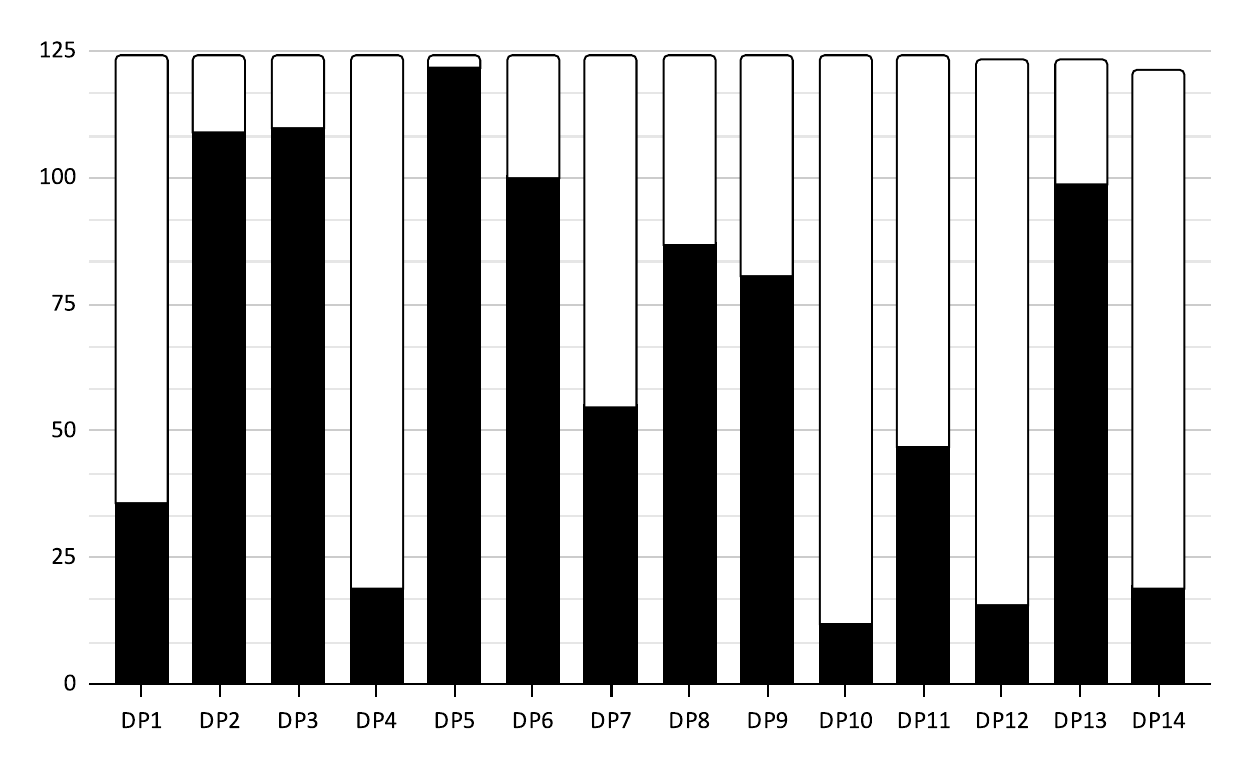}
    \includegraphics[width = .45
    \linewidth]{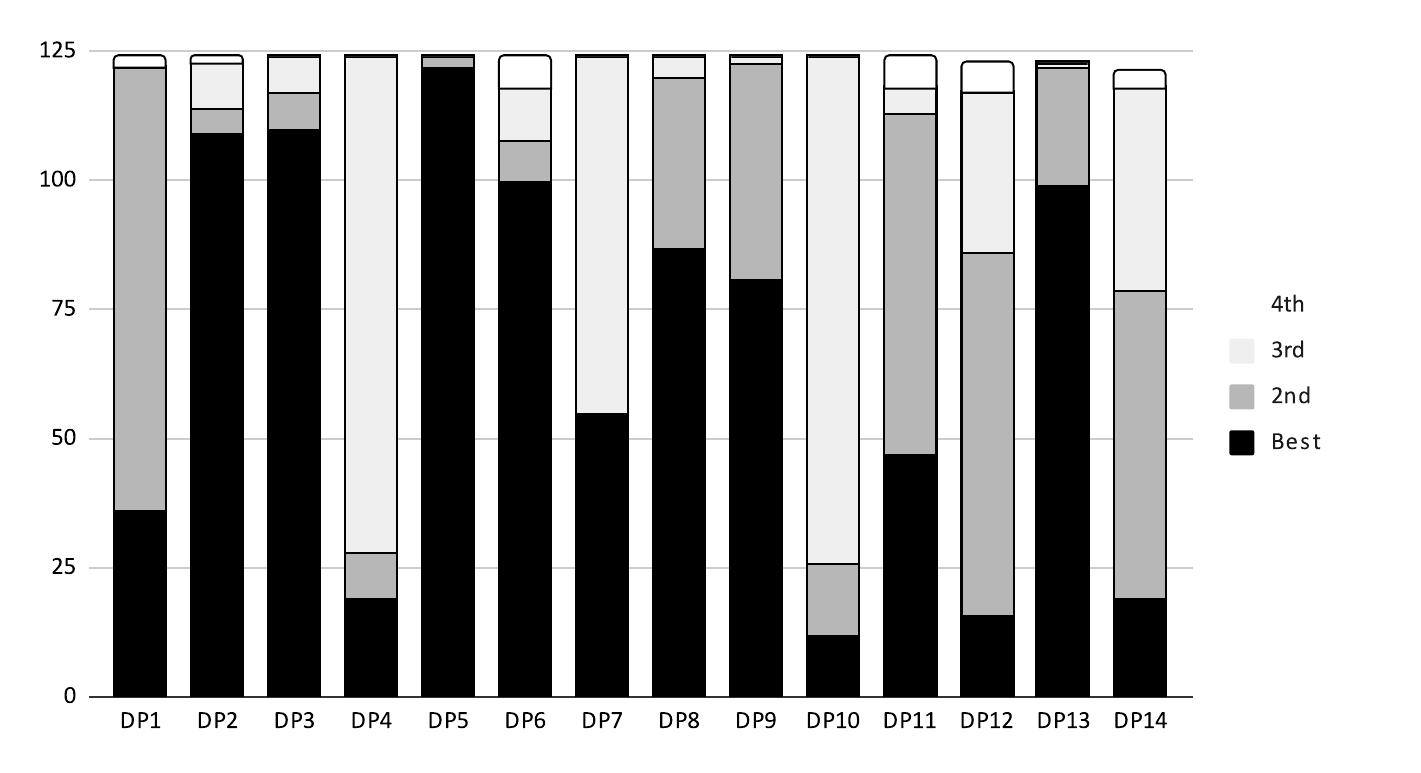}
    \caption{Illustration of the visual difference between the binary prediction framing vs one of the partial credit systems we propose applied to reanalyze Study 2's data from the Four Towers domain (\cite{Anderson2019, anderson2020mental}).
    \textbf{Left}: Overview of \emph{binary} prediction correctness for every participant, showing correct predictions in black and incorrect predictions in white.
    Both floor and ceiling effects are prevalent, which hinder comparative statistics.
    \textbf{Right}: Overview of distributions of loss in rank $LR()$ for every participant at every prediction.
    Note that even with only 4 choices, participants still very rarely picked some options.
    }
    \label{fig4TowersOverview}
\end{figure}

Anderson et al.~\cite{Anderson2019, anderson2020mental} already provided an analysis using the binary prediction framing, reporting no significant results based on the prediction data.
Figure~\ref{fig4TowersOverview} provides a high level look at that data by combining all treatments and focusing on the decisions (Anderson et al.~\cite{anderson2020mental}'s Figure 10 provides the data split by treatment, so we will omit that data for brevity).
Suffice to say, floor and ceiling effects abound, once again.

\subsection{Adopting the proposed framing, $LV()$ and $LR()$}

For value space, we worked on the Q-values associated with each action taken at every decision point, which in this domain represent a prediction of the number of points the agent thinks it will obtain if it selects a particular action and then follows the policy.
Since we have 4 different quadrants that the AI can decide amongst, for the rank approach we did not do any discretization ($DLR()$, Equation~\ref{eqnDLR}).
Over all 14 decisions, we found that the Q values ranged from -366 to 53.
We also noticed that for many DPs, the gap between two consecutively ranked Q-values was large.
For example, the Q-values for DP1 were (in rank order) 
31, -28, -284, -313.
In this case, the difference between the top choice and second best choice is 59, whereas the difference between second and third-ranked actions is 256---a gap five times greater.

\begin{figure}
    \centering
    \includegraphics[width = 0.475\linewidth]{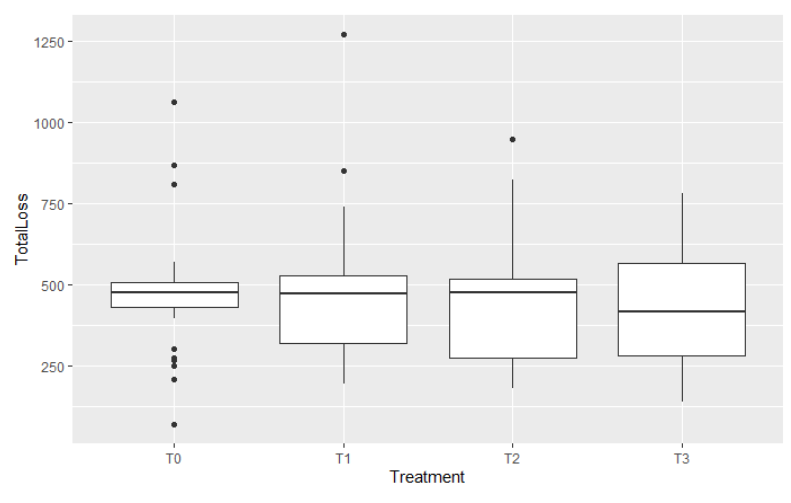}
    \includegraphics[width = 0.475\linewidth]{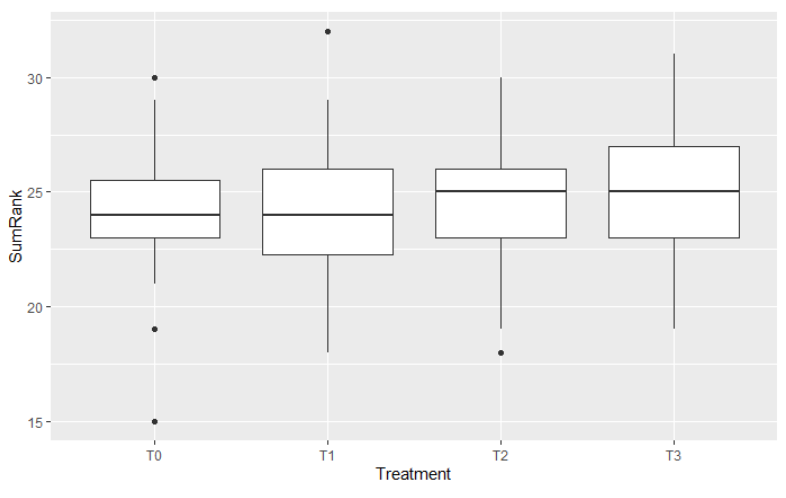}
    \caption{\textbf{Left}: Distributions of the sum of loss in value ($LV()$, y-axis) based on all predictions from all participants in each treatment (x-axis).
    The values range from -366 to 53 per decision point and the aggregate sum of loss goes up to 1250.
    For each treatment type, there are around 30 participants.
    \textbf{Right}: Distributions of the sum of loss in rank ($LR$, y-axis) based on predictions from participants in each treatment (x-axis).
    We have 4 quadrants and they are ranked from 1 to 4. 
    Similar to value space, we have around 30 participants in each treatment and the aggregate sum ranges from 0 to 38.
    The higher the participants' rank/value loss from the median line, the worse they performed, meaning up is bad on both charts. }
    \label{figure4TowersBoxPlot}
\end{figure}

In cases like this, we might expect $LV()$ (visualized in Figure~\ref{figure4TowersBoxPlot}) to outperform $LR()$, but alas, we do not find significant differences in either of our framings during our re-analysis.

\boldify{Checking normality and variances}

First, we tested the normality of participants' prediction scores, revealing that only one distribution exhibited normality, the one based on rank space (Shapiro-Wilk, W = 0.98844, p=0.3997).
In contrast, participants' data based on value space does not exhibit normality (Shapiro-Wilk, W = 0.93836, p=3.088e-05).

Subsequently, we checked if the participants' data showed equal variances among each other.
The results indicated that both the participants' prediction loss in value space (Levene's Test, df=3, f-value=0.1237 p=0.9459) and rank space (Levene's Test, df=3, f-value=0.2278, p=0.8769) met the requisite conditions for homogeneity of variances with median centering.

\boldify{We check.. now what? best alternative is KW test}

In light of the observed deviations from normality in two of the datasets, a Kruskal-Wallis test is an appropriate alternative to ANOVA to compare the prediction quality of participants in different treatments.
Unfortunately, we did not find any statistically significant differences between the median prediction losses between any pair of treatment groups in either rank space (Kruskal-Wallis, chi-squared = 1.894, df = 3, p=.5947) or value space (Kruskal-Wallis, chi-squared = 0.69606, df = 3, p=0.8741).

\boldify{Since we didn't find significance, we are moving to indicators. How we define them}

In the subsequent phase of our analysis, we embarked on an evaluation of feature importance.
In this endeavor, we identified a total of 18 binary indicator variables associated with each decision point.
These indicator variables encompass a variety of features describing a state, such as the presence or absence of a friendly tank in a given quadrant and the presence of enemy tanks, among others. 
Specifically, these variables were encoded as binary values, with a value of 1 signifying the presence of an enemy tank and 0 denoting its absence.

\boldify{How we select these variables.}

To ensure the appropriateness of retaining all these variables, we conducted a multicollinearity test, shown in Table~\ref{tableVIF}. 
The results revealed a noteworthy correlation between the values associated with the "BEFany" (big enemy fort) and ``ETany'' (enemy tank) variables and several other variables.
Based on this information we took BEFany out of the model.

\begin{table}
    \centering
    \begin{tabular}{@{}l| r r r@{}}
    & \textbf{GVIF}
    & \textbf{Df}
    & \textbf{GVIF\^{}(1/(2*Df))}
    \\\hline
    
    Treatment
    & 1.001694
    & 3
    & 1.000282
    \\

    BEFany
    & 36.738070
    & 1
    & 6.061194
    \\

    BFFany
    & 1.082859
    & 1
    & 1.040605
    \\

    SEFany
    & 1.339894
    & 1
    & 1.157538
    \\

    SFFany
    & 1.756536
    & 1
    & 1.325344
    \\

    BFCany
    & 4.805654
    & 1
    & 2.192180
    \\

    SFCany
    & 1.373462
    & 1
    & 1.171948
    \\

    ETany
    & 30.213202
    & 1
    & 5.496654
    \\

    \end{tabular}
   
    \caption{Variable Inflation Factor (VIF) test results, which is a measure of the amount of multicollinearity (correlation between independent variables) in regression analysis.
    Higher VIF suggests a stronger linear relationship between the independent variables.
    Here, our VIFs of 30 far exceed the typical threshold for concern, which is 10.}
    \label{tableVIF}
\end{table}

\boldify{Last step is to analyze the selected variables.}

Subsequently, we employed a backward stepwise regression procedure, guided by the Akaike Information Criterion (AIC), to systematically eliminate variables from our model. 
This iterative approach commenced with the full regression model containing all 18 indicator variables and no other terms.
Next, the process iteratively removed the variable that contributed the least to our model's predictive capacity. 
Unfortunately, the variable selection process removes treatment first, since it is unimportant to the model.

\boldify{what did we find.. tada}


\subsection{Adopting the proposed framing, $mRBO$}

We then turn to our other proposed framing, the $mRBO$, whose results are visible in Table~\ref{tabFourTowersMonolithic}.
Unfortunately, this approach yields a lot of ties in columns, which makes it much harder to compare treatments.
The table shows participants in the $NONE$ treatment provided the best (or tied for best) votes 8 times, Saliency claimed the same number, Reward bars only 7, and Both at 10.
Notably, Anderson et al.~\cite{anderson2020mental}'s original analysis found that participants in the Both treatment tended to outperform the other groups as well.

\begin{table*}
    \centering
    \small
    \begin{tabular}{@{}l |llll | llll | lll | lll@{}}
    & \multicolumn{14}{c}{\textbf{$mRBO$}}
    \\
    \textbf{Treatment}
    & DP1 & DP2 & DP3 & DP4
    & DP5 & DP6 & DP7 & DP8 
    & DP9 & DP10 & DP11
    & DP12 & DP13 & DP14
    \\\hline\hline

    \NONE{}
    & \textbf{0.9}
    & 0.955
    & \textbf{1.0}
    & \textbf{0.855}

    & \textbf{1.0}
    & 0.685
    & 0.855
    & \textbf{1.0}
    
    & \textbf{1.0}
    & \textbf{0.855}
    & 0.73
    
    & 0.855
    & \textbf{1.0}
    & 0.855
    \\\hline

    Saliency Maps
    & 0.63
    & 0.955
    & 0.955
    & \textbf{0.855}

    & 0.55
    & \textbf{1.0}
    & 0.855
    & \textbf{1.0}
    
    & \textbf{1.0}
    & \textbf{0.855}
    & \textbf{0.9}
    
    & 0.855
    & \textbf{1.0}
    & \textbf{1.0}
    \\\hline

    Reward Bars
    & 0.63
    & 0.955
    & 0.955
    & \textbf{0.855}

    & 0.55
    & \textbf{1.0}
    & 0.855
    & \textbf{1.0}
    
    & \textbf{1.0}
    & \textbf{0.855}
    & \textbf{0.9}
    
    & 0.855
    & \textbf{1.0}
    & 0.855
    \\\hline

    Both
    & \textbf{0.9}
    & \textbf{1.0}
    & 0.955
    & \textbf{0.855}

    & \textbf{1.0}
    & 0.955
    & \textbf{0.955}
    & \textbf{1.0}
    
    & \textbf{1.0}
    & \textbf{0.855}
    & 0.63
    
    & \textbf{0.9}
    & \textbf{1.0}
    & 0.855
    \\\hline

    \end{tabular}
    \caption{Four Towers domain results from applying Equation~\ref{eqnMRBO} to each treatment group at each prediction and also combining all predictions.
    The best score in each column is bold, which reveals a lot of ties and similar counts of top-quality voting patterns from each treatment.}
    \label{tabFourTowersMonolithic}

\end{table*}

In summary, our analysis for the Four Towers domain consistently suggests a lack of discernible differences in predictive performance across various treatment levels, even after incorporating the indicator variables into the model.

\section{Discussion}
\label{secDiscussion}

To analyze the theory we propose in this paper, we will employ the framework by Sj{\o}berg et al.~\cite{sjoberg2008theories}, focusing in particular on Explanatory Power and Generalizability.

\subsection{RQ1 revisited: Were our analysis techniques good enough? About Explanatory Power}

This answer is unclear at this time.
Our approach does make sense and seems to provide better-looking visualizations.
However, the new lens did not reveal much significance beyond what we had been able to see with the binary framing since we basically moved from \emph{zero} significant differences to \emph{one} significant difference.
Floor and ceiling effects persist and cause a great deal of trouble.

The presence of floor/ceiling effects in our results raises  question about ``Explanatory Power'' (The degree to which a theory accounts for and predicts all known observations within its scope~\cite{sjoberg2008theories}).
In some cases, these effects can arise from one of two things:
1) a binary insistence on ``if it's not perfect, it's wrong'', giving no credit for near-optimal answers or
2) natural clustering of participants' responses.
Point 1 creates artificial floor/ceiling effects, and our methods will ameliorate them.
However, Point 2's floor/ceiling effects are ``real'', meaning that changes in measures cannot make them disappear, nor \textit{should} they.
We present two examples of these ``real'' effects in Figures~\ref{figMNKoverview} and \ref{figGraded}, when many participants made the exact same prediction:
in one case ``this move wins'', the prediction was easy, so the ceiling effect persisted, as it should.
In others, the prediction was harder (e.g., Predictions 2 and 3 featured ``this move makes 3-in-a-row''), and so the floor effect persisted, as it should since participants' responses were clustered.

What our methods do is handle artificial (overly conservative floor/ceilings), thereby differentiating treatments more effectively than binary prediction.

\subsection{In which contexts are these analysis techniques well defined? About Generalizability}
\label{sectionDiscGeneralize}

\boldify{In which contexts are these well-defined?}

The generality (has breadth of scope, independence of specific settings~\cite{sjoberg2008theories}) of our approach is sufficient that it works for tasks beyond sequential domain action selection.
In particular, it will work for regression, as well as classification, as long as the classification is based on class probabilities.
This is because in regression or classification with class probabilities, the learning problem looks more like Equation~\ref{eqnScores} or \ref{eqnScoresForall} than Equation~\ref{eqnOneAction}.

As mentioned in Section~\ref{secBackgroundHybrid}, the neural network this paper relies upon is tasked with learning the function described in Equation~\ref{eqnScoresForall}.
That design choice enables quick computing of loss, rankings, etc; meaning all of the analysis strategies we outlined would be well-defined.
That said, employing the function in Equation~\ref{eqnOneAction} constrains the system's output to a binary correct/incorrect framing.
On the other hand, employing Equation~\ref{eqnScores} opens up the system's outputs to any of the proposed strategies in this section, so long as it is not too computationally expensive to perform that function for \emph{enough} actions, for some definition of enough.
Executing the function for \emph{all} actions is required to compute a full rank ordering for Equations~\ref{eqnLR} ($LR$) and \ref{eqnDLR} ($DLR$).
Since it may be expensive to determine the agent's full rank ordering without doing so, very large action (or continuous) spaces may prefer utilizing a partial ordering or values instead of ranks via Equation~\ref{eqnLV} ($LV$).

We intend our methods\footnote{Note that our equations do not include the discretized version of Equation~\ref{eqnLV} ($LV$).
We built this equation but did not present it in the paper since it seemed unnecessary.
However, if we are wrong, the construction of $DLV$ would look similar to $DLR$ (Equation~\ref{eqnDLR}).} to scale to large output spaces, as per Section~\ref{sectionIntro}'s examples based on StarCraft and autonomous driving.
That said, only further studies on other domains can accurately determine generality.

\subsection{Why not incorporate speed to produce the prediction?}
\label{sectionDiscSpeed}

\boldify{Why not use speed? It punishes some folks, depending on their cognitive styles}

Previous research (e.g., \cite{soratana2021human, driggs2016communicating}) has occasionally used \emph{speed} of participants' response to help judge prediction quality.
We chose not to do this for several reasons.
First, we felt that including speed would punish participants with certain cognitive styles.
For example, depending on the amount of information provided, one might hypothesize that a \emph{selective} information processor might respond faster than a \emph{comprehensive} information processor (terminology from~\cite{burnett2016gendermag-jrnl}).
Second, for this kind of task, we want to encourage participants to engage system II thinking\footnote{%
\textit{``System 1 is intuitive thinking, often following mental shortcuts and heuristics; System 2 is analytical
thinking, relying on careful reasoning of information and arguments. ... while
XAI techniques make an implicit assumption that people can and will attend to every bit of explanations, in reality,
people are more likely to engage in System 1 thinking.''}~\cite{liao2022humancentered}
}.

\subsection{How many prediction choices are optimal?}

\boldify{We don't know, depends on your definition of similar}

The answer to this question is not yet clear.
Our methods are designed to account for near-optimal ``misses'', but a researcher could use domain knowledge to decide the level of ``nearness'' that they want to consider.
As an example, in the basketball domain, one might create an equivalence class for jump shots, which would combine 2-point and 3-point attempts.
Alternatively, the researcher might create an equivalence class for  2-point attempts, combining dunks and short jump shots despite their visual dissimilarity.
The main difference in framing between these two views is whether they define as similar actions that \textit{appear similar} (i.e., jump shots) or \textit{produce similar outcomes} (i.e., 2-point attempts).

\section{Threats to Validity}

As Wohlin et al.~\cite{Wohlin-2012} observe, every study has threats to validity.
As such, we employ Yin's framework~\cite{yinBook} when reporting on our biases and attempts to mitigate them.

\boldify{learning effects: each participant did all 3 tasks, seeing each explanation (possibly multiple times) with NNs encoding a similar policy.
We worked to mitigate this by randomizing the task order but maintaining the same results, in terms of moves the agents made and game results.
One advantage our design has is that each participant saw EACH task and EACH expl, meaning they could provide direct comparison if they so chose.
Unfortunately, this type of direct comparison was fairly rare and rarely separated the task and the expl}

While this paper only focused on the Prediction Task, each participant did other tasks too (Ranking and Comparison).
As a result of these extra tasks, there could be learning effects present in our study, both with respect to how agents behave and which explanations the participants preferred.
We worked to mitigate this threat by randomizing the task order.
Despite the random order, we seeded the random number generator so that each participant would see the same results; both in terms of game results and moves made within the game.
One of the goals with our design was to ensure each participant saw all tasks and all explanations, hoping for direct comparison/contrast (unfortunately rare).

\boldify{While we showed each participant multiple explanations, we do not have any guarantee that our participants ACTUALLY looked at or used them.
Eye tracking might help us to know what they saw, but does not offer the same degree of control as a randomized clinical trial, where the clinicians can ensure that the participant receives the treatment the protocol randomly selected for that person.}

Some treatments incorporated multiple explanations, but we had no way to guarantee each participants used, understood, or even looked at them.
This allowed participants some degree of choice in selecting their own treatment, as opposed to being bound by our random selection.
Eye tracking might help us know what they looked at, but would still not give much information about understanding or usage.

\boldify{Agents interpreted as random: Since we did fully not solve the domain and the mutant intervention introduced random behavior, some humans interpreted multiple agents as behaving randomly, as opposed to considering it in terms of ``degrees of randomness'' (or similar).
To mitigate this as best we could we tried to take the top performing mutants, though we needed quite a few to run the study (2x for tutorial, 2x for comparison task, 4x for ranking task, 2x for predict task? (maybe 4? I forget exactly)}

We did not fully solve the domain and the mutant agent generation~\cite{dodge2022people} approach introduces randomness.
Thus, participants seemed to occasionally decline to assess the {\color{cyan}Sky}-colored agent, thinking it was behaving randomly, e.g.:
\quotateInset{P61 \OTB{}}{012312052	1	ABI	TIM	TIM	TIM	TIM	(4, 1)	0	\OTB{}}
{(4, 1):::I have little reason to support this, other than it being a possible "good" move. F2 and D2 are also possibilities, given the {\color{cyan}Sky} has used the diagonals towards the other player before.}
{I have little reason to support this, other than it being a possible "good" move. F2 and D2 are also possibilities, given the {\color{cyan}Sky} has used the diagonals towards the other player before.}

\quotateInset{P78	\NONE{}}{012618071	3	TIM	ABI	TIM	TIM	ABI	(2, 2)	0	\NONE{}}
{(2, 2):::It would be a good move to block {\color{magenta}Magenta}, but I'm not feeling very confident in my choices right now. It's difficult to predict {\color{cyan}Sky} sometimes or get myself to think like agent {\color{cyan}Sky}}
{It would be a good move to block {\color{magenta}Magenta}, but I'm not feeling very confident in my choices right now. It's difficult to predict {\color{cyan}Sky} sometimes or get myself to think like agent {\color{cyan}Sky}}

While we selected the top performing mutants, we still needed around a dozen different agents to perform our study, each with differing degrees of randomness.

\boldify{low motivation to take tasks seriously: game obviously low stakes, participants holds no stake in the assessment (as, e.g., an IBM analyst would for a system IBM was about to deploy).}

Another reason participants might not try very hard in the assessment was that they were not a stakeholder.
Their compensation was not tied to their task performance, nor would the deployment of a poor quality agent reflect on them personally.
Fortunately, for these reasons, an industry XAI analyst has more motivation to assess rigorously.

\boldify{Sampling biases from AI-related studies: these kinds of high-tech projects sometimes attract more Tims in terms of efficacy and learning style than in the general population.
To mitigate this, the best we could do is use the sample median as described in analysis (occasionally creating an imbalanced split).
While we COULD aggregate across moves to improve our stat power, we did not because this whole study is about DISaggregation, both in terms of person performing the assessment and the situation the assessment covers.
}

\boldify{standard issue stuff about lab studies: outside participants' normal workflow, under observation, etc}

Our recruiting materials described an AI-related study, which could affect interested parties.
Further, since we conducted the study in-lab, the typical threats for lab studies apply.
These include participants being: outside normal workflow,
outside normal work environment, under observation, and
free from distractions, such as cell phones.

\section{Conclusion}

In this paper, we made three contributions: 
\begin{enumerate}
    \item Three new methodologies (loss in value, loss in rank, and modified rank-biased overlap) for analyzing data from human predictions of AI behavior;
    \item a new in-lab study with 86 participants;
    \item and a re-analysis of data from a previously published study with 124 participants~\cite{anderson2020mental,Anderson2019}.
\end{enumerate}
In both analyses, treating prediction as a binary (right or wrong) would have yielded \textit{zero} statistically significant results.
Meanwhile, our approach was able to improve that a little, obtaining a single statistically significant result.

The advantages of using our analysis methodologies are that it is able to:
\begin{itemize}
\item \textit{Reduce risk of Type 2 error} - 
The results we shoed in Section~\ref{secMNKlVlR} indicate that we did find at least one case where a previous analysis found \textit{``no differences... between study groups when, in fact, there was''}~\cite{shreffler2023}.
We accomplish this effect by differentiating between degrees of wrongness in the incorrect prediction.

\item \textit{Apply to many contexts} -
As we argue in Section~\ref{sectionDiscGeneralize}, the strategies outlined in this paper apply to many regression or multi-class classification problems.

\item \textit{Reduce the size of the prediction space} -
In Section~\ref{sectionDiscretize}, we introduced how a discretization function can shrink the action space down by creating equivalence classes among actions.
This is a flexible strategy to potentially decrease the complexity of the problem for participants.
\end{itemize}

The drawbacks of using our approach are that it may not:
\begin{itemize}

\item \textit{Reduce risk of Type 2 error as much as another analysis strategy} -
We view it as important to not let good be the enemy of perfect.
We showed how some risk of Type 2 error is inherent due to clustering of participant responses creating floor and ceiling effects, while other risk is reducible via measurement choice.
Hopefully future work will more successfully mitigate the effects, where possible.


\item \textit{Help analyze binary predictions} -
We view this as not much of a problem, since traditional notions of classification accuracy apply in this setting.

\item \textit{Include participant certainty or response speed currently} -
As we argue in Section~\ref{sectionDiscSpeed}, response speed seems like an unreliable element to incorporate into the measure.
Certainty seems much more valuable, but we leave it to future work to combine our methods with those Bondi et al~\cite{bondi2022role}proposed for incorporating certainty. 
\end{itemize}

While it remains to be seen how severely the floor and ceiling effects continue to affect XAI researchers, using our approach can add shades of gray to data that appears black-and-white at first.

\bibliographystyle{ACM-Reference-Format}
\bibliography{00-paper}

\newpage
\pagestyle{empty}
\end{document}